\documentclass[aps,prd,showkeys,nofootinbib,superscriptaddress,tightenlines,11pt]{revtex4-2}

\usepackage[a-2u]{pdfx}
\usepackage[utf8]{inputenc}
\usepackage{lmodern}

\usepackage{amsmath}        
\usepackage{amsfonts}       
\usepackage{dsfont}			
\usepackage{amsthm}         
\usepackage{bbding}         
\usepackage{bm}             
\usepackage{graphicx}       
\usepackage{fancyvrb}       
\usepackage{dcolumn}        
\usepackage{booktabs}       
\usepackage{paralist}       
\usepackage{xcolor}         
\usepackage{physics}		
\usepackage{booktabs}
\usepackage{multirow}
\usepackage{accents}
\usepackage[normalem]{ulem}
\usepackage{hyphenat}
\usepackage{float}        
\usepackage{tikz}
\usetikzlibrary{intersections}
\usepackage{pgfplots}
\usepgfplotslibrary{fillbetween}
\usepackage{subcaption}
\hypersetup{unicode}
\hypersetup{breaklinks=true}

\pgfplotsset{compat=1.18}

\theoremstyle{plain}

\newcommand{\R}{\mathbb{R}}
\newcommand{\N}{\mathbb{N}}
\newcommand{\Z}{\mathbb{Z}}
\newcommand{\CC}{\mathbb{C}}

\DeclareMathOperator{\sgn}{sgn}

\newcommand{\goto}{\rightarrow}

\newcommand{\weyl}{\xrightarrow{\text{Weyl}}}

\newcommand{\ucharles}{Faculty of Mathematics and Physics, Charles University, V Hole\v{s}ovi\v{c}k\'{a}ch 2, 18000 Prague 8, Czech Republic}

\begin{document}

\title{Trace and diffeomorphism anomalies of the classical Liouville theory, \\
Virasoro algebras, Weyl-gauging and all that}

\author{Pavel Haman}
\email{pavel.haman@mff.cuni.cz}
\affiliation{\ucharles}
\author{Alfredo Iorio}
\email{alfredo.iorio@mff.cuni.cz}
\affiliation{\ucharles}

\date{\today}

\begin{abstract}
To fully clarify the invariance of the classical Liouville field theory under the Virasoro algebra, we first elucidate in detail the concept of \textit{classical anomaly}, discuss the occurrence of two symmetry algebras associated to this problem, and provide some new formulae to compute the classical center in a general fashion. We apply this to the study of the symmetries of the free boson in two dimensions. Moving to Liouville, we see how this gives rise to an energy\hyp{}momentum tensor with non-tensorial conformal transformations, in flat space, and a non-vanishing trace, in curved space. We provide a variety of improvements of the (local) theory, that restore Weyl invariance. With explicit computations, we show that the covariant conservation of the Weyl-invariance-improved energy\hyp{}momentum tensor is lost, in general, and relate the chosen improvement with the corresponding subset of preserved diffeomorphisms. The non-tensorial transformation rule of the Weyl-invariance-improved energy\hyp{}momentum tensor in curved space is explicitly back-traced to the Virasoro center.
\end{abstract}


\maketitle


\section{Introduction}

In this paper we expand and deepen the scope of the investigation reported in \cite{Letter1HamanIorio2023}, give a detailed explanation of many of the results reported there and provide new general results and discussions. Although the focus is still on Liouville field theory, what we say here can pave the way to further investigations in different models and with different scopes than ours here.

Liouville field theory is an exactly solvable two-dimensional model that enjoys a prominent role in many fields of the theoretical and mathematical investigation.

The story started in the 1850s, when Joseph Liouville posed and solved a problem of the  differential geometry of two-dimensional surfaces of constant Gaussian curvature, $K$, both positive and negative \citep{Liouville_1853}. That is to know the conformal factor of the metric, say $g_{\mu \nu}(x,y) = \delta_{\mu \nu} e^{2 \rho(x,y)}$. The equation that $\rho$ has to satisfy is \citep{Liouville_1853}
\begin{equation}
  \triangle \rho + K e^{2\rho} = 0 \,,
\end{equation}
with $\triangle = \partial_x^2 + \partial_y^2$.

The story kept going on and on, to the point that Liouville equation is nowadays ubiquitous in mathematical physics. We find it in lower dimensional quantum field theories, see, e.g., \citep{Jackiw1982ClassQuantLiouville}, in condensed matter models, see, e.g., \citep{Iorio2011WeylGaugeGraphene,iorio2020vortex}, and, most notably, in the early models of two-dimensional (quantum) gravity, see, e.g., \citep{QuantumGravity_Christensen}, governed by
\begin{equation}\label{eq:2Dgravity}
  R - \Lambda = 0 \,.
\end{equation}
Indeed, this equation, written in isothermal coordinates, turns out to be exactly Liouville equation\footnote{The Ricci scalar $R$ in two dimensions is related to $K$ as $R = 2 K$.}
\begin{equation}
  \Box \rho + \frac{\Lambda}{2}e^{2\rho} = 0 \,.
\end{equation}

Perhaps the most known field of application in mathematical physics, though, is string theory. From the early days of Polyakov's work \citep{Polyakov_1981}, till the nowadays very intense activity of the AdS/CFT correspondence \citep{Maldacena1999-gz}. Indeed, as well known, turning the point of view the other way around, Liouville theory is one of the conformal field theories corresponding to the three-dimensional gravity \citep{TasiOnAdsCft2016}.

It is then important to know its symmetries in all details, already at the classical level. In particular, Liouville theory is known to be a model enjoying both scale and full (global) conformal symmetries in flat space, hence it belongs to the cases studied in \citep{WeylGauging}. There the reasons why and when scale invariance of the classical theory (for fields of arbitrary spin in arbitrary dimensions, including Liouville) implies full conformal invariance are explained. A crucial ingredient of the recipe is the assumption that Weyl and diffemorphism invariances of such theories may hold together. Even though full (global) conformal symmetry is known to be in place, a more recent work \citep{WeylvsLiouville} makes the conjecture that Liouville theory might not be made both diffeomorphic and Weyl invariant, evoking a generic ``classical anomaly'' as the reason for that.

In \citep{Letter1HamanIorio2023} Jackiw's conjecture is proved and the solutions of those issues are presented. Here we study the problem from its basic building blocks, starting by identifying a precise mathematical definition of the suggestive idea of classical anomaly evoked in \citep{WeylvsLiouville}. We do so, by first defining a classically anomalous theory as a theory with genuine central charge in the algebra of Noether charges, elucidating the key role of the mismatch between the Hamiltonian and Lagrangian formulations, see, e.g., \citep{Toppan2001ClassicalAnomalies}. This way we clarify many ambiguities present in the literature and provide some new results, such as a way to generate the central charge in the case of spatiotemporal symmetries.

The focus here is on the gravitational anomalies, that we shall show to be present also at the classical level, and even in flat spacetime. Among those, Weyl or trace anomaly, found vast applications in \textit{quantized} theories embedded in curved spacetimes \citep{Christensen_1977}. There it is quantization that forces the breakdown of the invariance in the choice of regularization, and the best one can do is to keep the other invariance, i.e., the one under diffeomorphisms \citep{QFTCurvedSpace}. This is a spacetime instance of the phenomenon discovered in the early days of the Standard Model of particle physics, known as the axial or chiral anomaly \citep{Adler_1969, BellJackiw_1969}. Back then, these were seen as mere obstructions in the construction of consistent theories \citep{IntroQFT_Peskin, QFTandSM_Schwartz}. The modern approach, though, appreciates that anomalies are related to fluxes of the energy\hyp{}momentum, hence could be at the core of important phenomena, such as \textit{Hawking radiation}. This is the case of the two-dimensionial Weyl/trace anomaly \citep{Christensen_1977}, and of the breaking of diffeomorphism invariance in higher dimensions \citep{Wilczek_2005} which, together with the requirement of cancellation of the gauge anomaly and the existence of an event horizon, necessarily leads to a non-vanishing Hawking flux.

Besides the physical side of the story just recalled, mathematical advances in geometry and representation theory \citep{MathIntroCFT_Schottenloher,Jackiw_1985Cocycles} link anomalies with \textit{central extensions} of the algebra of symmetry generators, hence with an exact symmetry that is softly broken, in a very specific way. Although such breaking is usually of quantum origin, see, e.g. \citep{CFTDiFrancesco}, the algebra extensions are not exclusive to quantized theories and may emerge already at the classical level \citep{Toppan2001ClassicalAnomalies}.

The paper is organized as follows. To set up the notation, we start off with Noether theorem in Section II, to then move to symmetries in Hamiltonian formalism in Section III. This discussion will help clarifying, in Section IV, that the origin of classical anomalies is the mismatch between Lagrangian and Hamiltonian formulations. In Section V conformal transformations in two dimensions are recalled in the light of the previous Sections.
In Section VI we show that an affine center is there already for the free scalar in two dimensions, it can be set to zero, but it is genuinely there.
This paves the way to Section VII, where Liouville theory in flat space is shown to have a genuine center, that this time cannot be set to zero. The Section is closed by remarks of the general validity of some results. This anomalous conformal symmetry in flat space becomes important, for our analysis, on curved backgrounds, as shown in Sections VIII and IX. In the crucial Section X, the energy-momentum tensor (EMT) is studied in detail. The breaking of diffeomorphism invariance, that is among the most important results of this paper, is presented in Section XI. Last Section is devoted to our conclusions and five Appendices provide more details.


\section{Symmetries in Lagrangian formalism and Noether theorem} \label{Sec:Noether}

A \textit{classical} dynamical system has an \textit{anomalous symmetry} when it enjoys both a Hamiltonian and a Lagrangian description, but the algebra obeyed by the symmetry generators is a \textit{centrally extended} version of the commutator algebra obeyed by the symmetry transformations \citep{Toppan2001ClassicalAnomalies}.

The first algebra uses Poisson brackets, for which an Hamiltonian description is necessary, while the symmetry generators are those of Noether theorem, descending from the Lagrangian formulation. It is then the subtle discrepancies between the Lagrangian and the Hamiltonian descriptions at the core of the phenomenon of the classical anomaly. For the sake of clarifying these issues and to provide the notation that we shall later use, let us now start with the Noether theorem \citep{Banados2016Noether}.

Consider the action
\begin{equation}
  A[\Phi^i] = \int \dd[d]{x} \mathcal{L}(\Phi^i(x),\partial_\mu\Phi^i(x)) \,,
\end{equation}
with $\Phi^i(x)$ a generic field, or set of fields, and $\mathcal{L}$ the Lagrangian density, assumed to depend only on fields and their first derivatives. The latter, although not necessary, simplifies calculations and it is general enough.

A \emph{symmetry} is a transformation $\delta \Phi^i(x)$ such that
\begin{equation}\label{eq:ActionSymmetry}
  \delta A[\Phi^i,\delta \Phi^i] = \int \dd[d]{x}
  \dd_\mu\qty[K^\mu(\Phi^i(x), \delta \Phi^i(x))] \,,
\end{equation}
holds for every reasonable input $\Phi^i(x)$. Here $K^\mu$ is some functional of $\Phi^i(x)$, of $\delta \Phi^i$ and of their derivatives, and $\dd_\mu$ denotes a total derivative wrt $x^\mu$. Such transformations are denoted as $\delta_s \Phi^i(x)$. The symmetries we consider here belong to \emph{Lie groups}.

Varying the action, first by fixing the transformation and changing the inputs, and then by fixing the input and changing the transformations, finally fixing both, the inputs and the transformations, and comparing the two variations, one gets
\begin{equation}
  \delta A[\Phi^i,\delta_s \Phi^i]
  = \int \dd[d]{x} \dd_\mu\qty[K^\mu(\Phi^i(x),\delta_s\Phi^i(x))]
  \approx \int \dd[d]{x}
  \dd_\mu\qty(\pdv{\mathcal{L}}{(\partial_\mu\Phi^i)}\delta_s\Phi^i) \,,
\end{equation}
where the use of the Euler-Lagrange equations, $\dd_\mu\pdv{\mathcal{L}}{(\partial_\mu\Phi^i)}  - \pdv{\mathcal{L}}{\Phi^i} = 0$, is indicated by ``$\approx$''.

\emph{Noether theorem} then states that if an action has an off-shell symmetry, then there is an on-shell conserved quantity, $\dd_\mu J^\mu \approx 0$, where
\begin{equation}\label{eq:NoetherCurrent}
  J^\mu \equiv  \pdv{\mathcal{L}}{(\partial_\mu\Phi^i)}\delta_s\Phi^i - K^\mu \,,
\end{equation}
is the \emph{Noether current}. In other words, the \textit{Noether charge}
\begin{equation}\label{eq:NoetherCharge}
    Q[\Phi^i,\dot\Phi{}^i](t) = \int_V \dd[d-1]{x} J^0\,,
\end{equation}
is conserved, $\dv*{Q(t)}{t} \approx 0$.

The general form of the Noether current \eqref{eq:NoetherCurrent} applies to spatiotemporal and internal symmetries alike. It can be seen that $K^\mu$ is nonzero in the first case, and zero in the second. For the interesting case of supersymmetry, see \cite{iorio2000note,lor2000noether,iorio2001supersymmetric}

As for spatiotemporal symmetries, we take the {\it active transformations} point of view, hence will compare the original field and the transported field at the same point\footnote{More details on this are in the Appendix \ref{app:NoetherTheorem}.}. The natural operation to consider is then the \emph{Lie derivative} $(\mathcal L_f \Phi)(p)$, which is the measure of how much a tensor field changes along the flow generated by $f^\mu$ at a given spacetime point $p$. In other words, our spatiotemporal transformation of a generic functional $F[\Phi^i]$ is given by $\delta_s F[\Phi^i]\equiv
  \epsilon \mathcal L_f F[\Phi^i]
  \equiv
  F[\Phi^i+\epsilon\mathcal L_f\Phi^i] - F[\Phi^i]
  =
  F[\Phi'^i] - F[\Phi^i]$.

Lie derivatives evaluate all objects at the same point, hence commute with ordinary derivatives, see, e.g., \citep{GeometryTopologyPhysics_Nakahara}. Spatiotemporal symmetries of special interest are the isometries $\mathcal L_fg_{\mu\nu} = 0$ that, for Minkowski spacetime, give for $f^\mu$ the Poincaré group. The isometries equation can be recast into the covariant \emph{Killing equation}
\begin{equation}\label{eq:KillingEquation}
  \nabla_\mu f_\nu + \nabla_\nu f_\mu = 0 \,.
\end{equation}

Of special interest here are the conformal transformations, $g'_{\mu\nu}(x)=e^{2\lambda(x)} g_{\mu\nu}(x)$, that infinitesimally may be written as $\mathcal L_f g_{\mu\nu}(x) = 2\lambda(x) g_{\mu\nu}(x)$, where $\lambda$ is uniquely given by tracing, yielding the covariant \emph{conformal Killing equation}
\begin{equation}\label{eq:ConformalKilling}
  \nabla_\mu f_\nu + \nabla_\nu f_\mu = \frac{2}{d}g_{\mu\nu}\nabla_\rho f^\rho \,.
\end{equation}
In flat spacetime the solutions form the conformal group \citep{CFTDiFrancesco, DiverseTopics}.

Internal transformations only act on the fields, thus the passive interpretation based on the transport of the observer does not apply here, see Appendix \ref{app:NoetherTheorem} and \citep{GaugeTheory}. In accordance with spatiotemporal transformations, a local variation of the field can be introduced
$\Delta \Phi^i(x)\equiv \lim_{\epsilon\goto0}\frac1\epsilon\qty[{\Phi'_\epsilon}{}^i(x)-\Phi^i(x)]$ and this too commutes with ordinary derivatives, hence it fits our approach to Noether theorem.

Weyl transformation, important here, is a spatiotemporal transfromation that can be recast as internal as
\begin{equation}\label{eq:WeylTransformationMetric}
  g_{\mu\nu}(x)\goto g'_{\mu\nu}(x) = e^{2\omega(x)}g_{\mu\nu}(x)\,,
\end{equation}
and other fields get a scaling factor
\begin{equation}\label{eq:WeylTransformationField}
  \Phi^i(x)\goto\Phi'^i(x)=e^{d_{\Phi^i}\omega(x)}\Phi^i(x)\,,
\end{equation}
where $d_{\Phi^i}$ is the scaling dimension of the field. Conformal and Weyl transformations are similar and often connected, see, e.g., \citep{WeylGauging}, but they are not the same for their different origin and meaning. Nonetheless, the names ``conformal'', ``Weyl'' and, most of all, ``scale'' transformations are too often, if not mistakenly, used interchangeably in the literature, as told, e.g., in \citep{ScalevsConformal}.

~

With this, we can now safely treat both types of transformations, spatiotemporal and internal, in a unified manner, through the \emph{geometric transformation} of \citep{DiverseTopics}. For any infinitesimal transformation we have
\begin{equation}\label{eq:GeometricTransformation}
  \delta_f\Phi^i(x) \equiv \Phi'^i(x) - \Phi^i(x)
  =\begin{cases}
    \epsilon \mathcal L_f\Phi^i & \text{spatiotemporal,}
    \\
    \epsilon\Delta_f \Phi^i & \text{internal,}
  \end{cases}
\end{equation}
where $f$ represents any characterization of the transformation (e.g. a vector field $f^\mu$ for spatiotemporal flows) and $\epsilon$ is an infinitesimal parameter.

Having a spatiotemporal symmetry transformation characterized by a vector field $f^\mu$, the Noether current can be simplified into a Bessel-Hagen form, \citep{BesselHagen_1921}, defining the EMT\footnote{With $\Theta_{\mu\nu}$ we indicate the EMT in flat space,
while $T_{\mu\nu}$ is for curved space.} $\Theta_{\mu\nu}$
\begin{equation}\label{eq:BesselHagenForm}
  J_f^\mu = \Theta^{\mu\nu}f_\nu\,.
\end{equation}
For a translations, $K^\mu$ in \eqref{eq:NoetherCurrent} is $K^\mu = f^\mu \mathcal L$, which defines the \emph{canonical EMT}
\begin{equation}\label{eq:CanonicalEMTensor}
  {\Theta_{\text{can}}^{\mu\nu}}= \sum_i\pdv{\mathcal L}{(\partial_\mu\Phi^i)}\partial^\nu\Phi^i - \eta^{\mu\nu}\mathcal L \,,
\end{equation}
that, in general, is not symmetric nor traceless, thus \emph{improvements}, such as
\begin{equation}\label{eq:NoetherCurrentImprovement}
  K^\mu\goto K^\mu + \dd_\nu V^{\mu\nu} \,,
\end{equation}
where $V^{\mu\nu}=- V^{\nu\mu}$, are necessary to make it so. The Belinfante tensor is a well-known example \citep{DiverseTopics}.

For diffeomorphism invariant theories, the EMT can be derived in a variational way
\begin{equation}\label{eq:CurvedEMTensor}
  T^{\mu\nu}=-\frac{2}{\sqrt{-g}}\fdv{\mathcal{A}}{g_{\mu\nu}} \,,
\end{equation}
it is inherently symmetric and, in the flat limit, it gives the improved version of the canonical EMT of \eqref{eq:CanonicalEMTensor}.

With this procedure the invariances of the theory are immediately inherited by the EMT, so no improvements are necessary. Indeed,
\begin{equation}\label{eq:EMTensorInvariances}
   \delta_s \mathcal A \approx -\frac{1}{2}\int \dd[d]{x}\sqrt{-g} T^{\mu\nu}\delta_s g_{\mu\nu} = 0 \,,
\end{equation}
and for diffeomorphism invariance,
$\delta_s g_{\mu\nu} = \mathcal L_f g_{\mu\nu}$, it gives $\int \dd[d]{x}\sqrt{-g} f_\nu\nabla_\mu T^{\mu\nu} = 0$, where $\nabla_\mu$ is the covariant derivative. Since the above equation holds for any $f^\mu$, the diffeomorphism invariance implies the conservation $\nabla_\mu T^{\mu\nu} = 0$. Similarly, for Weyl transformation \eqref{eq:WeylTransformationMetric}, $\delta_s g_{\mu\nu} = 2\omega g_{\mu\nu}$, it gives
\begin{equation}\label{eq:WeylImpliesNoTrace}
  \int \dd[d]{x}\sqrt{-g} \omega T^{\mu}{}_\mu
  = 0\,,
\end{equation}
and the arbitrariness of $\omega(x)$ implies the vanishing trace of the EMT $T^\mu{}_\mu = 0$.

For infinitesimal transformations an important tool is the algebra of symmetry transformations. A commutator of two consecutive geometric transformations defines another geometric transformation
\begin{equation}\label{eq:AlgebraTransformations}
  [\delta_f,\delta_g]\Phi^i = \delta_h\Phi^i \,,
\end{equation}
with composition law $h^\mu = h^\mu(f, g)$.

For spatiotemporal symmetry transformations the algebra is given by commutators of Lie derivatives
$\comm{\mathcal L_g}{\mathcal L_f}\Phi^i = \mathcal L_\comm{g}{f} \Phi^i$ where the composition law
$\comm{g}{f}^\mu= g^\nu\partial_\nu f^\mu - f^\nu\partial_\nu g^\mu = \mathcal L_g f^\mu$ is the \emph{Lie bracket}. The generators of the transformations, $G_a$, can be introduced \citep{CFTDiFrancesco}
\begin{equation}\label{eq:TransformationGenerator}
  \delta \Phi^i(x)=\Phi'^i(x)-\Phi^i(x)\equiv\sum_a\epsilon^a G_a\Phi^i(x) \,,
\end{equation}
to simplify the algebra to
\begin{equation}\label{eq:LieAlgebraTransformations}
  \comm{G_a}{G_b}\Phi^i = {f_{ab}}^c G_c \Phi^i \,,
\end{equation}
where ${f_{ab}}^c$ are the structure constants of the Lie algebra.

\section{Symmetries in Hamiltonian formalism} \label{Sec:NoetherInverse}

A theory in Hamiltonian mechanics is characterized by its Hamiltonian $H$ and the Poisson brackets operator $\pb{\cdot}{\cdot}$ built upon a phase space. Then, the transformations of phase-space variables, or associated functionals, are ruled by the Poisson brackets operator and by the appropriate generator (e.g., the Hamiltonian $H$ is the generator of the time translations). For us, the Hamiltonian density, $\mathcal H$, depends only on fields, $\Phi^i$, conjugated momenta, $\Pi_i$, and spatial derivatives $\grad$ of the fields
\begin{equation}
  H[\Phi^i, \Pi_i](t) = \int\dd[d-1]{\vb{x}}\mathcal{H}(\Phi^i(\vb{x},t), \Pi_i(\vb{x},t),\grad\Phi^i(\vb{x},t)).
\end{equation}
For non-singular Lagrangians that condition descends from the Legendre transformation, while for singular Lagrangians we take this condition as a reasonable assumption, as explained in Appendix \ref{app:SingularLagrangian}. With the usual definition of Poisson brackets between two local functionals, for a field and its conjugated momentum we have
\begin{equation}
\pb{\Phi^i(\vb{x},t)}{\Pi_j(\vb{y},t)} = \delta^i_j\delta(\vb{x}-\vb{y}) \,.
\end{equation}

The dynamics of a generic quantity $F[\Phi^i,\Pi_i](t)$ is governed by
\begin{equation}\label{eq:HamiltonianDynamics}
  \dv{F[\Phi^i,\Pi_i](t)}{t} \approx \pb{F(t)}{H(t)} + \pdv{F[\Phi^i,\Pi_i](t)}{t}\,,
\end{equation}
where $\approx$ denote the use of Hamilton equations.

In this Section, we want to prove a somewhat ``inverse statement'' of the Noether theorem, namely that a conserved charge generates a symmetry transformation via Poisson brackets.
More precisely, if the transformation $\delta_f\Phi^i$ is a symmetry of an action $A$ and $Q[\Phi^i, \Pi_i | f](t)$ is the associated Noether charge, then
\begin{equation}\label{eq:NoetherChargeAsGenerator}
    \delta_f\Phi^i = \pb{\Phi^i}{Q[f]} \,,
\end{equation}
which is stated in \emph{Hamiltonian formalism} while the Noether charge is obtained in \emph{Lagrangian formalism}. Therefore, the proof requires reformulating quantities in Hamiltonian version. It is assumed there exists an invertible relation between conjugated momenta $\Pi_i$ and time derivative of the fields $\dot\Phi{}^i$. The only quantity containing time derivatives of the fields is the \emph{Hamiltonian Lagrangian density}
$\mathcal{L}_H(\Phi^i, \dot\Phi{}^i,\grad\Phi^i,\Pi_i) = \dot\Phi{}^i\Pi_i - \mathcal{H}(\Phi^i,\grad\Phi^i,\Pi_i)$.

Before moving to the proof of \eqref{eq:NoetherChargeAsGenerator} we derive a useful off-shell identity for conserved quantities. Noether charge $Q[f]$ satisfies $\dv*{Q[f]}{t}\approx 0$, but on-shell $Q[f]$ also satisfies \eqref{eq:HamiltonianDynamics}, $\pb{Q[f]}{H} + \pdv*{Q[f]}{t} \approx 0$. Since there are no time derivatives of the fields, the equations of motion are not actually needed, hence if that expression is zero on-shell, it is so off-shell
\begin{equation}\label{eq:ConservationOffShell}
  \pb{Q[f]}{H}+\pdv{Q[f]}{t} = 0 \,.
\end{equation}

With this, we can now prove that the Noether charge is a symmetry generator. We shall compute the total time derivative of the charge, through Poisson brackets, and then taking the time derivative of its definition. By comparing the two results we should be able to see the result.

The total time derivative of a generic functional $F$ is
\begin{equation}
    \dv{t}F[\Phi^i,\Pi_i](t)=\int\dd[d-1]{\vb{x}}\qty(\fdv{F}{\Phi^i}\dot\Phi{}^i+\fdv{F}{\Pi_i}\dot\Pi_i)+\pdv{t}F[\Phi^i,\Pi_i](t) \,.
\end{equation}
If $F$ is the Noether charge, hence it satisfies \eqref{eq:ConservationOffShell}, and
\begin{equation}\label{eq:TotalDerivativeFunctional}
    \dv{}{t}Q[\Phi^i, \Pi_i | f](t) = \int\dd[d-1]{\vb{x}}\qty[\fdv{Q}{\Phi^i}\qty(\dot\Phi{}^i-\fdv{H}{\Pi_i})+\fdv{Q}{\Pi_i}\qty(\dot\Pi_i+\fdv{H}{\Phi^i})] \,.
\end{equation}

To compute the total time derivative of $Q[f]$ from its definition \eqref{eq:NoetherCharge}, the Hamiltonian version is taken as the starting point
\begin{equation}
  Q[\Phi^i, \Pi_i | f](t) = \int\dd[d-1]{\vb{x}}J^0 = \int\dd[d-1]{\vb{x}}\qty(\pdv{\mathcal{L}_H}{(\partial_0\Phi^i)}\delta_f\Phi^i-K_H^0) \,,
\end{equation}
hence
\begin{equation}\label{eq:TotalDerivativeNoetherCharge}
        \dv{}{t}Q[\Phi^i, \Pi_i | f](t) = \int\dd[d-1]{\vb{x}}\qty[-\delta_f\Pi_i\qty(\dot\Phi{}^i-\fdv{H}{\Pi_i})+\delta_f\Phi^i\qty(\dot\Pi_i+\fdv{H}{\Phi^i})].
\end{equation}
Comparing \eqref{eq:TotalDerivativeNoetherCharge} to \eqref{eq:TotalDerivativeFunctional}, and taking into account that both equations are off-shell, hence the fields $\Pi_i$ and $\Phi^i$ are arbitrary and independent, the quantities in the parentheses are arbitrary, non-zero and independent. Therefore
\begin{equation}
    \begin{split}
        \delta_f\Pi_i =& -\fdv{Q}{\Phi^i} \quad\Rightarrow\quad \delta_f\Pi_i= \pb{\Pi_i}{Q[f]} \,,
        \\
        \delta_f\Phi^i =&~ \fdv{Q}{\Pi_i} \phantom{-}\quad\Rightarrow\quad\delta_f\Phi^i= \pb{\Phi^i}{Q[f]} \,.
    \end{split}
\end{equation}
This proves \eqref{eq:NoetherChargeAsGenerator}.

Here we assumed smooth boundary conditions, which may not always be ensured. In case of more general asymptotic behavior and boundary te
rms, it is often necessary to use equations of motion. Nonetheless, in the general case, the Noether charges always generate symmetry transformations \emph{on-shell}.

\section{Classical anomalies, a tale of two algebras} \label{sec:ClassicalAnomalies}

Although, the study of centrally extended algebras of symmetries are of great interest in quantum physics, there has been only a handful of studies on the central extension at the classical level, \citep{Brown1986CentralCharges}. The strong emphasis on quantized theories might induce to believe that central charges are quantum effects only. The scope of this work is to show how a center appears in a very important classical model, Liouville theory, and to study the physical implications of these on \textit{classical anomalies}.

Symmetries of a theory manifests themselves as an algebra of symmetry transformations, on the one hand, and as an algebra of Noether charges, on the other. If they are manifestations of the same phenomenon, only written within different formalisms, these two algebras need to be isomorphic. This is what we want to explore here, and show how this is at the core of the phenomenon of classical anomalies \cite{Toppan2001ClassicalAnomalies}.

In the Lagrangian formalism, the infinitesimal symmetries manifest themselves in the algebra of transformations. It is only natural to expect an equivalent structure in Hamiltonian formalism. Indeed, it can be shown that Poisson brackets of two conserved charges is again a conserved quantity,
$\dv*{\pb{Q[f]}{Q[g]}}{t} \approx 0$, hence Noether charges form a closed algebra under Poisson brackets. Thus, in correspondence to \eqref{eq:AlgebraTransformations}, the algebra of charges can be written as
\begin{equation}\label{eq:AlgebraCharges}
  \pb{Q[f]}{Q[g]} = Q[k] \,,
\end{equation}
where $k^\mu=k^\mu(f,g)$ is the composition law. Assuming a basis of Noether charges $Q_a$ reduces the previous equation to
\begin{equation}\label{eq:LieAlgebraCharges}
  \pb{Q_a}{Q_b}={g_{ab}}^c Q_c \,,
\end{equation}
where ${g_{ab}}^c$ are the structure constants of the Lie algebra of charges. It can be proved that, if $Q_a$ and $Q_b$ are well\hyp{}defined, then the result of their Poisson brackets is well\hyp{}defined too \citep{Brown1986PoissonBrackets}.

As shown in the previous Section, the transformations on the phase space can be generated via Noether charges, \eqref{eq:NoetherChargeAsGenerator}. A general object $F[\Phi^i, \Pi_i]$ transforms as
$\delta_f F[\Phi^i, \Pi_i] = \pb{F}{Q[f]}$. Applying to this $\delta_g$, leads to the question of how the $f^\mu$ transforms.

If $f^\mu$ does not transform, $\delta_g f^\mu = 0$, then $\delta_f$  defines a group of \textit{internal transformations} $\Delta_f$ on the phase space, hence $\Delta_g\Delta_f F[\Phi^i,\Pi_i] = \pb{\pb{F}{Q[f]}}{Q[g]}$. Using the Jacobi identity for Poisson brackets, gives the wanted connection between algebras
\begin{equation}\label{eq:AlgebrasConnection}
  \comm{\Delta_g}{\Delta_f}F = \pb{F}{\pb{Q[f]}{Q[g]}} \,,
\end{equation}
and composition rules \eqref{eq:AlgebraTransformations}, $\comm{\Delta_g}{\Delta_f}F = \Delta_{h(g,f)}F$, and \eqref{eq:AlgebraCharges}, $\pb{Q[f]}{Q[g]} = Q[k(f,g)]$, given by
\begin{equation}
  h^\mu(g,f) = k^\mu(f,g)  \,.
\end{equation}
This can be seen from $\comm{\Delta_g}{\Delta_f}F=\Delta_hF=\pb{F}{Q[h]}$, compared with \eqref{eq:AlgebrasConnection}.

We found here that the two composition rules have opposite order or, equivalently, opposite sign $k^\mu(f,g) = - h^\mu(f, g)$, due to $h^\mu(f,g)=-h^\mu(g,f)$. It is true that one can always redefine the charges, $Q[f]\goto - Q[f]$, to have the composition law in the same order, that proves the isomorphism between the algebras. However, this redefinition does modify the transformation rule \eqref{eq:NoetherChargeAsGenerator}.

When we deal with \textit{spatiotemporal transformations}, $\delta_g f^\mu\neq 0$ mingles the spacetime upon which the phase space is built together with Poisson brackets. A rule for spatiotemporal transformations can be found by comparing the Lie derivative $\mathcal L_f$ with a specifically chosen internal transformation $\Delta_f$ with the following transformation rules
\begin{equation}\label{eq:InternalLieDerivative}
  \Delta_f\Phi^i = \mathcal L_f\Phi^i\,.
\end{equation}
Although they transform dynamical fields the same, the two differ significantly while acting on another non-dynamical spacetime vector field $g^\mu$
\begin{align}
  \Delta_fg^\mu =&~ 0\,, \label{eq:InternalActingOnVectorField}
  \\
  \mathcal{L}_fg^\mu =&~ f^\nu\partial_\nu g^\mu - g^\nu\partial_\nu f^\mu
  = \comm{f}{g}^\mu\,.\label{eq:LieBracket}
\end{align}
This allows to find the composition rule $h^\mu(g,f)$, for a generic field $\Phi^i$
\begin{align}
  \comm{\Delta_g}{\Delta_f}\Phi^i  = \Delta_{h(g,f)}\Phi^i \,,
\end{align}
with $h^\mu(g,f) = f^\nu\partial_\nu g^\mu - g^\nu\partial_\nu f^\mu = -h^\mu(f,g)=\mathcal L_f g^\mu$, the last equality coming from \eqref{eq:LieBracket}.

To find how the algebra of spatiotemporal transformations and the algebra of Noether charges are related, it is helpful to realize that the previous result implies
\begin{equation}\label{eq:SpacetimeInternalTransformationConnection}
  \mathcal L_f\mathcal L_g\Phi^i = \mathcal L_f\Delta_g\Phi^i = \Delta_f\Delta_g\Phi^i-\Delta_{h(f,g)}\Phi^i = \Delta_g\Delta_f\Phi^i\,,
\end{equation}
where the last equality was obtained by the use of the commutator of two transformations $\Delta$. Thus, two spatiotemporal transformations follows the opposite order to the ``internalized'' version of the transformations.

Since transformations $\Delta_f$ are generated in Hamiltonian mechanics via Poisson brackets, and since the equation \eqref{eq:SpacetimeInternalTransformationConnection} connects these and Lie derivatives $\mathcal L_f$, a rule for generating $\mathcal L_f$ without altering Poisson brackets under the action of the transformation can be derived. Clearly, $\comm{\mathcal L_g}{\mathcal L_f}\Phi = \comm{\Delta_f}{\Delta_g}\Phi = \Delta_{h(f,g)}\Phi$ and $\mathcal L_f\mathcal L_g \Phi = \pb{\pb{\Phi}{Q[f]}}{Q[g]}$.

The general rule can be deduced from these results. For transformations not changing the phase space the transformations are generated via \eqref{eq:NoetherChargeAsGenerator} and multiple transformations are adding up as
\begin{equation}
  \Delta_f\cdots\Delta_g\Delta_h\Phi = \pb{\pb{\pb{\pb{\Phi}{Q[h]}}{Q[g]}}{\cdots}}{Q[f]} \,.
\end{equation}
On the other hand, the spatiotemporal transformations may lead to additional structures or changes in the phase space. Thus, the final rule for generating spatiotemporal transformations is
\begin{equation}
  \mathcal L_f \mathcal L_g\cdots \mathcal L_h \Phi = \pb{\pb{\pb{\pb{\Phi}{Q[f]}}{Q[g]}}{\cdots}}{Q[h]} \,.
\end{equation}
The opposite order in the ladder of Noether charges for internal and spatiotemporal transformations should be noted.

The above shows how the composition rule of the algebra of Noether charges and the composition rule of algebra of transformations are connected
\begin{equation}\label{eq:AlgebrasConnectionGeneral}
  \comm{\Delta_f}{\Delta_g}\Phi = \pb{\Phi}{\pb{Q[g]}{Q[f]}}
  \qc
  \comm{\mathcal L_f}{\mathcal L_g}\Phi = \pb{\Phi}{\pb{Q[f]}{Q[g]}}\,.
\end{equation}
However, the algebra of charges can be centrally extended, still respecting the algebra of transformations the charges generate. Indeed, the Noether charges may obey
\begin{equation}\label{eq:AnomalousChargeAlgebra}
  \pb{Q[f]}{Q[g]} = Q[k] + K[f,g] \,,
\end{equation}
where $k$ is given by the composition rule as in \eqref{eq:AlgebraCharges} and $K$ is a quantity independent of fields and conjugated momenta, thus, it has vanishing Poisson brackets with any field
\begin{equation}\label{eq:CentralCharge}
  \pb{K[f,g]}{\Phi^i} = 0 \,.
\end{equation}
Moreover, $\pb{K[f,g]}{Q[h]} = 0$, for any $f,g,h$. This defines $K$ as the \emph{central charge} and it shows that the transformations generated through \eqref{eq:NoetherChargeAsGenerator} by $\pb{Q[f]}{Q[g]}$ and by $Q[k(f,g)]$ are identical, and both charge algebras, without the center \eqref{eq:AlgebraCharges} or with the center \eqref{eq:AnomalousChargeAlgebra},  give rise to the same algebra of transformations \eqref{eq:AlgebraTransformations}. Thus, the presence of a central charge is not reflected in the algebra of transformations.

Now, Noether charges are not unique, as shown by \eqref{eq:NoetherCurrentImprovement}, and this has an effect on central charges too, since a class of improved charges $Q'$ can always be found
\begin{equation}\label{eq:ChargeTrivialShift}
  Q'[f]= Q[f] + q[f],
\end{equation}
with $q$ being a field independent quantity. Although they both generate the same transformations, $Q'$ obeys a centrally extended version of the algebra obeyed by $Q$, \eqref{eq:AlgebraCharges}, $\pb{Q'[f]}{Q'[g]} = Q'[k(f,g)] - q[k(f,g)]$. Such object $q$ has vanishing Poisson brackets with any charge $Q'$, thus, it is a central charge. We see, then that redefining the charges, a central charge can be added or removed at will from the algebra, but this does not provide any new information. Such center is not a general functional of $f^\mu$ and $g^\mu$, rather a functional of their specific combination $k^\mu(f,g)$, the composition law of the algebra. As this procedure is invertible, any part of central charge proportional only to the $k$ is removable, by a simple redefinition of the charges. If the whole center is removable, it is called a \emph{trivial} central charge. If it cannot be removed, in the way just explained, it is a \emph{genuine} central charge.

Theories that require genuine extension of algebra of charges are called \emph{anomalous} \citep{Toppan2001ClassicalAnomalies}. In what follows we shall always remove trivial central charges, if not stated otherwise.

The central term appeared in the Hamiltonian reformulation of the theory, in the algebra of charges. Nonetheless, central term can be witnessed even in Lagrangian formulations by realizing that on-shell
$\Delta_gQ[f] \approx \pb{Q[f]}{Q[g]}$. Thus, the center may be studied already in the Lagrangian formalism, \citep{Blagojevic_2005}, by studying the transformations of charges
\begin{equation}\label{eq:LagrangianCenter}
  \Delta_g Q[f] \approx Q[h(g,f)] + K[f,g].
\end{equation}

To close this Section, let us show an intriguing property that holds for spatiotemporal transformations
\begin{equation}\label{NewResult1}
\mathcal L_g Q[f] \approx K[f,g] \,,
\end{equation}
and let us stress that, as a part of the algebra of charges, the central charge has to be conserved, i.e., $\pb{K}{H}+\pdv{K}{t} = 0$ implies $\pdv{K}{t} = 0$.

\section{Conformal transformations in two dimensions} \label{Sec:ConfTwoDim}

In a two-dimensional flat spacetime the conformal Killing equation is
\begin{equation}\label{eq:2DConfKilling}
  \partial_\mu f_\nu + \partial_\nu f_\mu = \eta_{\mu\nu}\partial_\rho f^\rho \,,
\end{equation}
with $\eta_{\mu\nu}$ the flat metric in Euclidean or Minkowski signature, see Appendix \ref{app:EuclideMinkowski}. Taking a derivative, this becomes $\partial_\nu\partial^\nu f_\mu = 0$, that is the wave equation, in Minkowski signature, or the Laplace equation, in Euclidean signature.

The focus here is on two-dimensional scalar field theories\footnote{Classical anomalies are not restricted to scalar theories, though. See \citep{Toppan2001ClassicalAnomalies} and \citep{Banados2016Noether}.}. The geometric conformal transformation of a scalar field $\phi(x)$ is
\begin{equation}\label{eq:FieldConformalTransformation}
  \delta_f\phi(x) = \phi'(x)-\phi(x) = \epsilon f^\mu\partial_\mu\phi\,.
\end{equation}

For the Euclidean plane, the solutions to the conformal Killing equation are (anti)holomorphic functions $f$ ($\bar f$). Hence, they can be expanded as a Laurent series
\begin{equation}\label{eq:EuclideanBasis}
  f(z) = \sum_{n\in\mathds{Z}} f_n z^n,\qquad \bar f(\bar z) = \sum_{n\in\mathds{Z}} \bar f_n \bar z^n\,.
\end{equation}

For the Minkowski spacetime the solutions are the right-going, $f^-$, and the left-going, $f^+$, waves. Assuming the $P$\hyp{}periodicity, see Appendix \ref{app:EuclideMinkowski}, the waves can be expanded into a Fourier basis
\begin{equation}\label{eq:MinkowskianBasis}
   f^\pm(x^\pm) = a_0 +\sum_{k\in\N}\qty[a_k\cos(\frac{2\pi}{P}kx^\pm)+b_k\sin(\frac{2\pi}{P}kx^\pm)] = \sum_{n\in \Z}f^\pm_n e^{i\frac{2\pi}{P}nx^\pm} \,.
\end{equation}

These allow to introduce the generators as (see \eqref{eq:TransformationGenerator})
\begin{equation}\label{lnEuclide}
l_n = -z^{n+1}\partial, \qquad \bar l_n = -\bar z^{n+1}\bar \partial \,,
\end{equation}
in Euclidean (complex) spacetime, and as
\begin{equation}\label{lnMinkowski}
 l_n^\pm \equiv \frac{iP}{2\pi} e^{i\frac{2\pi}{P}nx^\pm}\partial_\pm \,,
\end{equation}
in Minkowski spacetime, where $i P/(2\pi)$ is a normalization factor.

Computing the commutators one gets the \textit{Witt algebras} \citep{MathIntroCFT_Schottenloher} in both notations
\begin{equation}\label{WittEuclide}
  \comm{l_n}{l_m} = (n-m)\,l_{n+m}  \,\qc \comm{\bar l_n}{\bar l_m} = (n-m)\,\bar l_{n+m}  \,\qc \comm{l_n}{\bar l_m} = 0 \,,
\end{equation}
and
\begin{equation}\label{WittMinkowski}
\comm{l_n^+}{l_m^+} = (n-m)\,l_{n+m}^+ \,\qc \comm{l_n^-}{l_m^-} = (n-m)\,l_{n+m}^-  \,\qc \comm{l_n^+}{l_m^-} = 0 \,,
\end{equation}
respectively, whose unique \textit{genuine} central extension are the \emph{Virasoro algebras} \citep{MathIntroCFT_Schottenloher}, e.g.,
\begin{align}\label{eq:VirasoroAlgebra}
  \comm{l_n^\pm}{l_m^\pm} =&~ (n-m)l_{n+m}^\pm + \frac{c}{12}n(n^2-1)\delta_{m+n,0} \,,
  \\
  \comm{l_n^\pm}{c} =&~ 0 \,,
\end{align}
for Minkowski spacetime, where $c$ is the central term, $1/12$ is a normalization factor.

We need not focus on other metrics than the Euclidean and Minkowski, because all two-dimensional spacetimes are \emph{locally conformally flat} \citep{Chern1955IsothermalCoordinates, GeometryTopologyPhysics_Nakahara}. Thus, for every two-dimensional (pseudo-)Riemannian manifold, \emph{isothermal coordinates} exist such that
\begin{equation}\label{eq:IsothermalCoordinates}
  \hat g_{\mu\nu}(x) = e^{2\rho(x)}\eta_{\mu\nu} \,,
\end{equation}
where the hat indicates quantities in the isothermal coordinate frame.

When a theory, on top of conformal invariance, also has Weyl invariance, a Weyl transformation suppresses the conformal factor, hence the computations can be restricted to flat metric spacetimes just discussed here and in Appendix \ref{app:EuclideMinkowski}. The connection between scale, conformal, diffeomorphism and Weyl invariances is explained in \citep{WeylGauging}, for classical theories in arbitrary dimensions, including two, and for fields of arbitrary spin. The case of the two dimensional scalar Liouville theory, though, became the focus of more attention in \citep{WeylvsLiouville}. It is the scope of the rest of this paper to investigate these issues in all details.


\section{The free scalar in two dimensions and its affine center} \label{Sec:FreeScalar}

Before diving into the Liouville theory, it is instructive to apply the previous results to a simpler model, the free scalar in two dimensions
\begin{equation}\label{eq:FreeScalarTheory}
  A_{\scriptscriptstyle{FS}} [\varphi] = \int\dd[2]{x} \frac{1}{2}\partial_\mu\varphi\partial^\mu\varphi \,,
\end{equation}
and Minkowski metric, $\eta_{\mu\nu}=\operatorname{diag}(1,-1)$. This gives as equation of motion
\begin{equation}\label{eq:FreeFieldEoM}
  \Box_{\text M}\varphi=\partial_\mu\partial^\mu\varphi=0\,.
\end{equation}

The theory is invariant under the infinitesimal conformal transformations \eqref{eq:FieldConformalTransformation} where $\epsilon$ is the infinitesimal parameter, and $f^\mu$ satisfies the conformal Killing equation \eqref{eq:ConformalKilling},
hence
\begin{equation}\label{eq:ConfKillingEqChapter2}
  \Box_{\text M} f_\mu = 0 \,.
\end{equation}

According to Noether theorem, there is a conserved current \eqref{eq:NoetherCurrent}, $J^\mu_f$, associated to the symmetry
\begin{equation}
  J^\mu_f = \qty(\partial^\mu\varphi\partial^\nu\varphi-\frac12\eta^{\mu\nu}\partial_\lambda\varphi\partial^\lambda\varphi)\epsilon f_\nu \,.
\end{equation}
From the Bessel-Hagen formula \eqref{eq:BesselHagenForm}, $J_f^\mu = \theta^{\mu\nu}f_\nu$, the canonical EMT of the free field is
\begin{equation}
  \theta^{\mu\nu} = \partial^\mu\varphi\partial^\nu\varphi-\frac12\eta^{\mu\nu}\partial_\lambda\varphi\partial^\lambda\varphi \,,
\end{equation}
and it is symmetric and traceless, as required.

Besides conformal transformations, the theory is symmetric under an affine transformation \citep{Jackiw1982ClassQuantLiouville}, that is a field shift,
\begin{equation}
\delta_{\chi}\varphi(x)=\epsilon\chi(x) \,,
\end{equation}
with associated current given by
\begin{equation}
J^\mu_\chi=\epsilon\qty(\partial^\mu\varphi\chi-\varphi\partial^\mu\chi) \,.
\end{equation}
For this to happen, the shift too must satisfy the wave equation
\begin{equation}\label{eq:AffineTransformationCondition}
  \Box_{\text M}\chi=0 \,.
\end{equation}

All the involved quantities, $\varphi$, $f^\mu$ and $\chi$, obey the wave equation, \eqref{eq:FreeFieldEoM}, \eqref{eq:ConfKillingEqChapter2}, \eqref{eq:AffineTransformationCondition}. Thus light-cone coordinates, discussed in Appendix \ref{app:EuclideMinkowski}, are the most natural: $\varphi(x) \approx \varphi^+(x^+) + \varphi^-(x^-)$, $\chi(x) = \chi^+(x^+) + \chi^-(x^-)$, and $f^\pm(x^\pm) = \frac1{\sqrt2}\qty(f^t(x)\pm f^x(x))$.

The derived quantities, currents $J^\mu_f$, $J^\mu_\chi$ and EMT $\theta^{\mu\nu}$, split to $x^+$ and $x^-$ dependent parts correspondingly, on-shell, under the natural assumption, $\delta_\chi\varphi^+(x^+) = \epsilon\chi^+(x^+)$,
$\delta_\chi\varphi^-(x^-) = \epsilon\chi^-(x^-)$.

Then it can be seen that $\theta_{++} = \partial_+\varphi\partial_+\varphi \approx \theta_{++}(x^+)$, $\theta_{+-} = 0$, $\theta_{--} = \partial_-\varphi\partial_-\varphi \approx \theta_{--}(x^-)$ and
${J_f}_\pm = \theta_{\pm\pm}f^\pm\approx {J_f}_\pm(x^\pm)$, ${J_\chi}_\pm \approx \partial_\pm\varphi^\pm\chi^\pm-\varphi^\pm\partial_\pm\chi^\pm  = {J_\chi}_\pm(x^\pm)$. Notice that, e.g., $J_+=J_+(x^+)$, but $J^+\neq J^+(x^+)$, becasue $J^+=J_-=J_-(x^-)$.

Therefore, the problem at hand splits in two independent parts, each with its own Noether charges: $\partial_\mu J^\mu = \partial_+ J_- + \partial_-J_+ \approx 0$, where each term vanishes independently. The charges can be defined as\footnote{Here we explicitly write the ``time'' variable. E.g., for the charge $Q^\pm$ the ``space'' variable, i.e., the integration variable of the definition \eqref{eq:NoetherCharge}, is $x^\pm$, while the other variable, $x^\mp$, plays the role of ``time''. This explicit notation is omitted in the following, but should be understood in the evaluation of ``equal-time'' Poisson brackets.}
\begin{align}
  Q^+[f](x^-) =& \int \dd{x^+} f^+(x^+)\theta_{++}(x^+)\,,
  \\
  Q^-[f](x^+) =& \int \dd{x^-} f^-(x^-)\theta_{--}(x^-)\,,
  \\
  Q[f](t) =&~ Q^+[f] + Q^-[f]\,,
\end{align}
for the transformation $\delta_f$ and
\begin{align}
  q^+[\chi](x^-) =&~ 2\int \dd{x^+}\chi^+\partial_+\varphi^+\,,
  \\
  q^-[\chi](x^+) =&~ 2\int \dd{x^-}\chi^-\partial_-\varphi^-\,,
  \\
  q[\chi](t) =&~ q^+[\chi] + q^-[\chi]\,,
\end{align}
for the transformation $\delta_\chi$.

The algebra of these charges can be only obtained by first constructing Poisson brackets in the light-cone Hamiltonian formalism. Since the Lagrangian is
\begin{equation}
  \mathcal{L} = \partial_+\varphi\partial_-\varphi\,,
\end{equation}
there are two obstacles. First, there is no natural time variable. Second, and more direct, the Lagrangian is linear in $\partial_\pm\varphi$, therefore the standard Legendre transformation fails to provide any meaningful Hamiltonian. Although, not knowing the Hamiltonian is not a complication for the reconstruction of the charge-algebra, not having a phase space is.
In Appendix \ref{app:SingularLagrangian} we discuss the issues and derive Poisson brackets for light-cone Lagrangian, which turn out to be
\begin{align}
  \eval{\pb{\varphi(x)}{\varphi(y)}}_{x^+=y^+} =&~
  \pb{\varphi^-(x^-)}{\varphi^-(y^-)} =
  -\frac14\sgn(x^--y^-) \,,
  \\
  \eval{\pb{\varphi(x)}{\varphi(y)}}_{x^-=y^-} =&~
  \pb{\varphi^+(x^+)}{\varphi^+(y^+)} =
  -\frac14\sgn(x^+-y^+) \,,
  \\
  \pb{A[\varphi]}{B[\varphi]} =& -\frac14\int \dd{z}\dd{z'}\sgn(z-z')\cdot
  \qty(
      \fdv{A[\varphi]}{\varphi^+(z)}\fdv{B[\varphi]}{\varphi^+(z')}
      + \fdv{A[\varphi]}{\varphi^-(z)}\fdv{B[\varphi]}{\varphi^-(z')}
  ) \,,
\end{align}
where $\sgn(x) = + 1 (-1)$ for $x > 0 ~(x < 0)$, and $\sgn(x) = 0$ for $x=0$, is the sign function. It can be equivalently rewritten in terms of Haeviside step function $\sgn(x) = 2H(x) - 1$, revealing a useful, distributive, property for calculations, $\sgn'(x) = 2 \delta(x)$. The provided pair of Poisson brackets (equal $x^+$ brackets and equal $x^-$ brackets), displays the split and separate evolutions along the $x^+$ and $x^-$ lines.

Having Poisson brackets it can be verified that the charges generate transformations as in \eqref{eq:NoetherChargeAsGenerator}, $\delta_f\Phi^i = \pb{\Phi^i}{Q[f]}$. With the use of the previous results and derivative of Poisson brackets,
$\pb{\varphi^\pm(x^\pm)}{\partial_\pm\varphi^\pm(y^\pm)} = \frac12\delta(x^\pm-y^\pm)$,
the transformations can be derived
\begin{align}
  \pb{\varphi^+(x^+)}{Q^+[f]} =&~ f^+(x^+)\partial_+\varphi^+(x^+) \,,\label{eq:SFChargesGenerator+}
  \\
  \pb{\varphi^-(x^-)}{Q^-[f]} =&~ f^-(x^-)\partial_-\varphi^-(x^-) \,,\label{eq:SFChargesGenerator-}
  \\
  \pb{\varphi^+(x^+)}{q^+[\chi]} =&~ \chi^+(x^+) \,,
  \\
  \pb{\varphi^-(x^-)}{q^-[\chi]} =&~ \chi^-(x^-) \,.
\end{align}

We can now construct the algebra of charges. In the following, conformal and affine charges are discussed separately, to emphasize their specific properties. At the end of the Section both algebras are brought together to reconstruct the Virasoro algebra.

The conformal charges $Q[f]$ close the conformal algebra
\begin{align}
  \pb{Q^\pm[f]}{Q^\pm[g]} =&~ Q^\pm[k]\,,
  \\
  \pb{Q^\pm[f]}{Q^\mp[g]} =&~ 0\,,
\end{align}
where $k^\mu$ is the Lie bracket \eqref{eq:LieBracket} of $f^\mu$ and $g^\mu$, $k^\mu = f^\nu\partial_\nu g^\mu - g^\nu\partial_\nu f^\mu$.

By expanding $f^\pm$ into Fourier series \eqref{eq:MinkowskianBasis} the conformal charges can be decomposed as follows
\begin{equation*}
  Q^\pm[f] = \int\dd{x^\pm} \theta_{\pm\pm}f^\pm
  = \sum_{n\in \Z} f^\pm_n \int\dd{x^\pm}\theta_{\pm\pm}e^{i\frac{2\pi}{P}nx^\pm}
  = -\frac{2\pi i}{P}\sum _{n\in \Z} f^\pm_n Q^\pm_n\,,
\end{equation*}
where
\begin{equation}\label{eq:LnAndTheta}
  Q^\pm_n\equiv \frac{P}{2\pi} \int\dd{x^\pm}\theta_{\pm\pm}e^{i\frac{2\pi}{P}nx^\pm}\,,
\end{equation}
form a basis of conformal charges. The factor $P/2\pi$ in the definition of $Q^\pm_n$ is chosen so that the confromal charges give the Witt algebra
\begin{align}\label{eq:ChoiceOfWittAlgebra}
  i\pb{Q^\pm_n}{Q^\pm_m}=&~(n-m)Q^\pm_{n+m}\,,
  \\
  i\pb{Q^\pm_n}{Q^\mp_m}=&~0 \,.
\end{align}
The presence of complex unit $i$ should not be over-looked. Nonetheless, the choice of having $i$ here is to obtain a non-imaginary central charge, as explained later. One can obtain the Witt algebra \eqref{WittMinkowski} by a redefinition $Q^\pm_n\goto -iQ^\pm_n$. Aspects of such redefinition are mentioned in Section \ref{sec:ClassicalAnomalies} and alter the connection between conformal charges \eqref{eq:LnAndTheta} and generators \eqref{lnMinkowski}
\begin{equation}
  l_n^\pm\varphi=\pb{\varphi}{Q^\pm_n}\,,
\end{equation}
as seen from \eqref{eq:SFChargesGenerator+} and \eqref{eq:SFChargesGenerator-}.

The definition \eqref{eq:LnAndTheta} shows that the charges $Q^\pm_n$ are Fourier modes of the EMT with suitable normalization\footnote{In the radial parametrization approach (see Appendix \ref{app:EuclideMinkowski}), the charges are the coefficients of the Laurent series of the EMT.}. When we sum all of them up, the full information about the EMT is recovered
\begin{equation}
  \theta_{\pm\pm}=\sum_{n\in\Z} \frac{2\pi}{P^2}Q^\pm_{-n}e^{i\frac{2\pi}{P}nx^\pm} \,.
\end{equation}

The affine charges $q[\chi]$, on the other hand, close a centrally extended algebra
\begin{align}
  \pb{q^\pm[\chi]}{q^\pm[\zeta]} =&~ K^\pm[\chi,\zeta] \,,
  \label{eq:FieldShiftCenter}
  \\
  \pb{q^\pm[\chi]}{q^\mp[\zeta]} =&~ 0 \,,
\end{align}
where
\begin{equation}
  K^\pm[\chi,\zeta] = \int\dd{x^\pm}
  \qty[
    \chi^\pm(x^\pm)\partial_\pm\zeta^\pm(x^\pm)
    - \zeta^\pm(x^\pm)\partial_\pm\chi^\pm(x^\pm)
  ] \,.
\end{equation}

The central charge $K$ introduced by the bracket \eqref{eq:FieldShiftCenter} is a field independent object, therefore, $\pb{\cdot}{K^\pm[\chi,\zeta]} = 0$, and, in particular, $\pb{\varphi}{K^\pm[\chi,\zeta]} = 0$. Thus, $K$ itself does not generate any transformation. The waves $\chi$ and $\zeta$ can be expanded into Fourier modes too
\begin{equation*}
  \chi^\pm(x^\pm) = \sum_{n\in\Z}\chi^\pm_n e^{i\frac{2\pi}{P}nx^\pm}
  \,,\qquad
  \zeta^\pm (x^\pm) = \sum_{n\in\Z}\zeta^\pm_n e^{i\frac{2\pi}{P}nx^\pm} \,,
\end{equation*}
leading to
\begin{equation}\label{eq:AffineCentralTerm}
  K^\pm[\chi,\zeta] = \sum_{n,m\in\Z} \chi^\pm_n\zeta^\pm_m\,2\pi i(m-n)\delta_{n+m,0} \,.
\end{equation}

Introducing a basis for affine generators $q^\pm_n \equiv q^\pm\qty[e^{i\frac{2\pi}{P}n x}]$ the equation \eqref{eq:FieldShiftCenter} can be written as
\begin{equation}
  i\pb{q^\pm_n}{q^\pm_m} = 4\pi n\,\delta_{n+m,0} \,,
\end{equation}
that is nothing else than a representation of the \emph{Heisenberg algebra}, \citep{CFTDiFrancesco},
\begin{equation}
  \comm{a_n}{a_m}=n\,\delta_{n+m,0}\,.
\end{equation}

The Poisson brackets between the charges close
\begin{align}
  \pb{Q^\pm[f]}{q^\pm[\chi]} =&~ q^\pm[f\cdot\partial\chi] \,,
  \label{eq:QqClosureness}
  \\
  \pb{Q^\pm[f]}{q^\mp[\chi]} =&~ 0 \,.
\end{align}
Thus a combined transformation can be chosen. Let us consider the following choice for $\chi$
\begin{equation}\label{eq:KillingFieldShift}
  \chi \equiv \partial\cdot f
  \quad\iff\quad
  \chi^\pm_n=f^\pm_n\frac{2\pi i}{P}n\,,
\end{equation}
that is available becasue $f^\mu$ itself is a solution of the wave equation. Now we can combine the two transformations, $\delta_f$ and $\delta_\chi$, parametrized only by $f^\mu$, into a new improved transformation, $\tilde\delta_f$, in the following way
\begin{equation}\label{eq:FreeImprovedConformalTransformation}
  \tilde\delta_f\varphi
  \equiv \delta_f \varphi + \frac1\gamma \, \delta_{\partial f}\varphi
  = f^\mu\partial_\mu \varphi + \frac1\gamma \, \partial_\mu f^\mu,
\end{equation}
where $\gamma$ is an arbitrary parameter. 

The role of a nonzero $1/\gamma$ here is that of merging the two independent transformations, $\delta_f$ and $\delta_\chi$, into one. In the rest of this Section we show the effects of this merging on the symmetry algebra of the free boson. In this case, given the arbitrariness of $\gamma$, such merging is unnecessary. On the other hand, in the next Section we shall demostrate that for other two-dimensional scalar theories the merging is inevitable, as it is the only way to have conformal symmetry.

The conserved current associated to the new transformation $\tilde\delta_f$ is
\begin{equation}
  \tilde J{}^\mu = J^\mu_f + \frac{1}{\gamma}J^\mu_{\partial f} \,,
\end{equation}
and it can be recast in the Bessel-Hagen form \eqref{eq:BesselHagenForm}, $\tilde J^\mu = \tilde\theta^{\mu\nu}f_\nu$, by improving the current $\tilde J^\mu$ by a superpotential-like term \eqref{eq:NoetherCurrentImprovement}
\begin{equation}
  \partial_\nu V^{\mu\nu} = 2\partial_\nu(f^\mu\partial^\nu\varphi-f^\nu\partial^\mu\varphi)
  +\partial_\nu(\varphi\partial^\mu f^\nu-\varphi\partial^\nu f^\mu) \,.
\end{equation}
By using \eqref{eq:ConformalKilling}, it can be shown that $J^\mu_{\partial f} + \partial_\nu V^{\mu\nu} = 2 f_\nu\qty(\eta^{\mu\nu}\Box_{\text M}-\partial^\mu\partial^\nu)\varphi$. Thus, the new EMT is
\begin{equation}
  \tilde\theta_{\mu\nu} = \partial_\mu\varphi\partial_\nu\varphi -\frac{1}{2}\eta_{\mu\nu}
  \partial_\alpha\varphi\partial^\alpha\varphi
  + \frac{2}{\gamma}(\eta_{\mu\nu}\Box_{\text M}-\partial_\mu\partial_\nu)\varphi\,.
\end{equation}

This new improved EMT $\tilde\theta_{\mu\nu}$ is not identically traceless, but it is so only on-shell. However, the split into $x^\pm$ parts is not violated,
\begin{align}
  \tilde\theta_{\pm\pm} =&~ \partial_\pm\varphi\partial_\pm\varphi -\frac2\gamma\partial_\pm^2\varphi\approx \tilde\theta_{\pm\pm}(x^\pm)\,,
  \\
  \tilde\theta_{+-} =&~\frac4\gamma\Box_{\text M}\varphi\approx 0\,,
\end{align}
and the new improved charges
\begin{equation}
    \tilde Q_\gamma^\pm[f]
    = \int \dd{x^\pm} \tilde\theta_{\pm\pm}f^\pm(x^\pm)
    = \int \dd{x^\pm}
    \qty[
        \partial_\pm\varphi^\pm\partial_\pm\varphi^\pm
        - \frac{2}{\gamma}\partial_\pm^2\varphi^\pm
        ]f^\pm
    = Q^\pm[f] +\frac{1}{\gamma}q^\pm[\partial\cdot f] \,,
\end{equation}
are still well defined.

From the previous results, the algebra of improved charges immediately follows
\begin{equation}\label{eq:ImprovedChargesAlgebra}
  \pb{\tilde Q_\gamma[f]}{\tilde Q_\gamma[g]} =
  \tilde Q_\gamma[k] +\frac{1}{\gamma^2}\Delta[f,g],
\end{equation}
where
\begin{equation}
  \Delta[f,g] = K[\partial f, \partial g]
  = \underbrace{\int \dd{x^+} (g^+\partial_+^3f^+- f^+\partial_+^3g^+)}_{\equiv\Delta^+[f,g]}
  + \underbrace{\int \dd{x^-} (g^-\partial_-^3f^--f^-\partial_-^3g^-)\,}_{\equiv\Delta^-[f,g]}.
\end{equation}

Combining \eqref{eq:AffineCentralTerm} and \eqref{eq:KillingFieldShift}, the central term immediately follows
\begin{equation}
  \Delta^\pm[f,g] = \sum_{n,m\in\Z} g^\pm_mf^\pm_n\qty(\frac{2\pi i}{P})^24\pi in^3\delta_{m+n,0}\,.
\end{equation}
Furthermore, the improved charges $\tilde Q^\pm_n$ can be introduced again as the Fourier decomposition of the improved EMT $\tilde\theta_{\mu\nu}$ with a proper normalization
\begin{equation}
  \tilde Q^\pm_n \equiv \frac{P}{2\pi}\int \dd{x^\pm} \tilde\theta_{\pm\pm} e^{i\frac{2\pi}{P}nx^\pm}\,.
\end{equation}
The centrally extended algebra \eqref{eq:ImprovedChargesAlgebra} can, thus, be recast as
\begin{equation}\label{eq:VirasoroImprovedFreeField}
  i\pb{\tilde Q_n^\pm}{\tilde Q_m^\pm} = (n-m)\tilde Q_{n+m}^\pm + \frac{4\pi }{\gamma^2}n^3\delta_{n+m,0}\,.
\end{equation}
Notice that the previously mentioned redefinition $Q^\pm_n\goto -iQ^\pm_n$ results in $c\goto -ic$.

The composition rule of the  Virasoro algebra \eqref{eq:VirasoroAlgebra}
\begin{equation*}
  \comm{l_n^\pm}{l_m^\pm} = (n-m)l_{n+m}^\pm + \frac{c}{12}n(n^2-1)\delta_{m+n,0} \,,
\end{equation*}
differs from \eqref{eq:VirasoroImprovedFreeField} by a term proportional to $n\,\delta_{n+m,0}=(n-m)/2\,\delta_{n+m,0}$. Nonetheless, it is easy to see, that it is a trivial (removable) center. First, note that for $k^\mu=f^\nu\partial_\nu g^\mu - g^\nu\partial_\nu f^\mu$ it holds
\begin{equation}
  \int \dd{x^\pm}k^\pm[f,g] = \sum_{n,m\in\Z}g^\pm_mf^\pm_n\, 2\pi i(m-n)\delta_{n+m,0}\,.
\end{equation}
Second, as indicated in \eqref{eq:ChargeTrivialShift}, we have the freedom to redefine $\tilde Q_\gamma[f]$ by adding terms proportional to $f$. A suitable choice is
\begin{equation}\label{eq:ImprovedChargeShift}
  \tilde Q_\gamma^\pm[f^\pm] \goto \tilde Q_\gamma^\pm[f^\pm] + \frac1{\gamma^2}\qty(\frac{2\pi i}{P})^2 \int\dd{x^\pm} f^\pm\,,
\end{equation}
leading to
\begin{equation}
  \Delta^\pm[f,g]\goto\Delta^\pm[f,g]-\qty(\frac{2\pi i}{P})^2\int\dd{x^\pm}k^\pm[f,g]
  = \sum_{n,m\in\Z} g^\pm_mf^\pm_n\qty(\frac{2\pi i}{P})^24\pi i n(n^2-1)\delta_{n+m,0}\,.
\end{equation}
This reconstructs the Virasoro algebra
\begin{equation}
  i\pb{\tilde Q_n^\pm}{\tilde Q_m^\pm} = (n-m)\tilde Q^\pm_{n+m} + \frac{4\pi}{\gamma^2}n(n^2-1)\delta_{n+m,0}\,,
\end{equation}
with a center
\begin{equation}\label{CenterFree}
  c = \frac{48\pi}{\gamma^2} \,.
\end{equation}

Let us conclude this Section with several comments. The conformal symmetry led to a simple Witt algebra, thus, the conformal symmetry of the classical free scalar is not anomalous. It was the introduction of the combined, improved, transformation, $\tilde \delta$, that gave rise to the genuine center and  resulted in the Virasoro algebra. Tracking the $1/\gamma$ factors suggests that the center of the Virasoro algebra is the affine center in new clothes, dressed upon the demand that the field shift has to be a conformal Killing field. Nonetheless, the theory itself does not restrict the choice of $\gamma$ in the definition and the choice $\gamma\goto\infty$ restores the anomaly free conformal transformations.

Although seemingly artificially constructed, the center of improved charges differs significantly from  the center of affine charges. While the value of the affine center can be changed by the renormalization of the affine charges $q^\pm_n$, the value of the Virasoro central charge cannot be normalized at will by the redefinition of the improved charges $\tilde Q_n^\pm$. This rigidity has physical consequences illustrated later.

In order to restore the Virasoro algebra with the usual composition law \eqref{eq:VirasoroAlgebra}, the improved charges had to be ``shifted''. The mathematical justification of trivial shifts of charges \eqref{eq:ChargeTrivialShift} was given in the previous Section and from its form \eqref{eq:ImprovedChargeShift} it can be seen that only the charge $\tilde Q_0$ changes its value in this case. Correspondingly, from the definition of charges
\begin{equation}
  \tilde Q_\gamma[f] \sim \int \tilde\theta\cdot f \,,
\end{equation}
the above shown shift can be understood as a shift of the EMT by a constant factor
\begin{equation}
  \tilde\theta_{\mu\nu} \goto \tilde\theta_{\mu\nu} -\frac{4\pi^2}{\gamma^2 P^2}\eta_{\mu\nu} \,.
\end{equation}
Such procedure thus corresponds to a shift of the mean value of the EMT.

\section{Liouville theory in flat space and its genuine center} \label{Sec:LiouvilleFlat}

Classical Liouville theory is genuinely conformally anomalous. In fact, contrary to the free scalar, Liouville theory
\begin{equation} \label{eq:FlatLiouvilleTheory}
  A_{\scriptscriptstyle{L}}[\Phi] = \int \dd[2]{x}
  \qty(
    \frac{1}{2}\partial_\mu\Phi\partial^\mu\Phi
    - \frac{m^2}{\beta^2}e^{\beta\Phi}) \,,
\end{equation}
with $\eta_{\mu\nu}=\operatorname{diag}(1,-1)$, with equation of motion
\begin{equation}\label{eq:FlatLiouvilleEoM}
  \Box_{\text M}\Phi + \frac{m^2}{\beta}e^{\beta\Phi} = 0 \,,
\end{equation}
is only symmetric under the \textit{improved} version of the conformal transformations.

Indeed, while the kinetic term is symmetric under standard conformal symmetry, as the free scalar, the potential term is not, because $\delta_f \qty(e^{\beta\Phi})  = f^\mu \partial_\mu \qty(e^{\beta\Phi})$ does not turn \eqref{eq:FlatLiouvilleTheory} into a pure divergence, as required by the Noether theorem to be a symmetry.

However, under the improvements extensively discussed earlier
\begin{equation}\label{eq:LiouvilleSymmetryTransformation}
  \tilde\delta_f \Phi = f^\mu\partial_\mu\Phi + \frac1\beta\partial_\mu f^\mu \,,
\end{equation}
we have $\tilde\delta_f \qty(e^{\beta\Phi}) = \partial_\mu \qty(f^\mu e^{\beta\Phi})$, and the theory is symmetric. 

Contrary to the free scalar, \eqref{eq:FreeImprovedConformalTransformation}, the parameter $\gamma$ is no longer arbitrary, as it is now fixed by the theory 
\[
\gamma = \beta \,.
\] 
The parameter $\beta$ is given, once and for all, in \eqref{eq:FlatLiouvilleTheory} and cannot be changed at will. In particular, one cannot set $\beta \to \infty$ or $\beta \to 0$, hence decouple the two transformations that make \eqref{eq:LiouvilleSymmetryTransformation}. Indeed, the theory is only invariant under the combination of the two, not under either transformation alone.

Moreover, the previously derived EMT only gets one extra contribution from the potential term
\begin{equation}\label{eq:FlatLiouvilleEMTimproved}
  \tilde\Theta_{\mu\nu} = \partial_\mu\Phi\partial_\nu\Phi -\eta_{\mu\nu}\qty(
    \frac{1}{2}\partial_\alpha\Phi\partial^\alpha\Phi - \frac{m^2}{\beta^2}e^{\beta\Phi})
  + \frac{2}{\beta}(\eta_{\mu\nu}\Box_{\text M}-\partial_\mu\partial_\nu)\Phi \,,
\end{equation}
and it is symmetric and traceless on-shell
\begin{equation}\label{eq:FlatLiouvilleEMTensor}
  \tilde\Theta^\mu_\mu = \frac{2}{\beta}\qty(\Box_{\text M}\Phi+\frac{m^2}{\beta}e^{\beta\Phi})\approx 0 \,.
\end{equation}
While the field $\Phi$ does not split on-shell into $\Phi^\pm$, as is the case for the free scalar, the EMT still does. By taking a derivative, it can be shown that $\partial_\mp\tilde\Theta_{\pm\pm} \approx 0$ and tracelessness implies $\tilde\Theta_{+-}\approx 0$. The conformal Killing vector $f^\mu$ also splits, thus meaningful Noether charges can be defined in the light-cone formulation
\begin{equation}\label{eq:LiouvilleNoetherCharge}
  \tilde Q^\pm[f](x^\mp) = \int \dd{x^\pm}\tilde \Theta_{\pm\pm}f^\pm =\int \dd{x^\pm} \qty((\partial_\pm\Phi)^2-\frac2\beta\partial^2_\pm\Phi)f^\pm \,.
\end{equation}

Given that the potential term does not interfere with the definition of conjugated momentum, Poisson brackets have the same form as for the free scalar
\begin{align}
  \eval{\pb{\Phi(x)}{\Phi(y)}}_{x^+=y^+} =&
  -\frac14\sgn(x^--y^-) \,,
  \\
  \eval{\pb{\Phi(x)}{\Phi(y)}}_{x^-=y^-} =&
  -\frac14\sgn(x^+-y^+) \,.
\end{align}

Again, it can be verified that the charges correctly generate the transformations
\begin{equation}
  \pb{\Phi(x^+,x^-)}{\tilde Q^\pm[f](x^\mp)}
  =
  f^\pm(x^\pm)\partial_\pm \Phi(x^+,x^-)
  + \frac1\beta \partial_\pm f^\pm(x^\pm)\,,
\end{equation}
and, from the previous discussion of the algebra of charges, it immediately follows
\begin{equation}\label{eq:LiouvilleChargeAlgebra}
  \pb{\tilde Q^\pm[f]}{\tilde Q^\pm[g]} =
  \tilde Q^\pm[k] + \frac{1}{\beta^2}\Delta^\pm[f,g] \,,
\end{equation}
where $k^\mu = f^\nu\partial_\nu g^\mu- g^\nu\partial_\nu f^\mu$ and
$\Delta^\pm[f,g] = \int \dd{x^\pm} (g^\pm\partial_\pm^3f^\pm- f^\pm\partial_\pm^3g^\pm)$.

Introducing the generators
\begin{equation}\label{eq:LiouvilleGenerator}
  \tilde Q^\pm_n \equiv \frac{P}{2\pi}\int \dd{x^\pm} \tilde\Theta_{\pm\pm} e^{i\frac{2\pi}{P}nx^\pm}= \frac{P}{2\pi}\tilde Q^\pm[e^{i\frac{2\pi}{P}nx^\pm}] \,,
\end{equation}
the algebra \eqref{eq:LiouvilleChargeAlgebra} can be recast into
\begin{equation}\label{eq:LiouvilleVirasoroAlgebra}
  i\pb{\tilde Q^\pm_n}{\tilde Q^\pm_m} = (n-m)\tilde Q^\pm_{n+m} + \frac{4\pi}{\beta^2}n^3\delta_{n+m,0} \,.
\end{equation}
Comparing this to the Virasoro algebra \eqref{eq:VirasoroAlgebra} gives the central charge for the classical Liouville theory
\begin{equation}\label{eq:CentralChargeLiouville}
  c = \frac{48\pi}{\beta^2} \,.
\end{equation}
This time the center is genuine since the symmetry of Liouville field, \eqref{eq:LiouvilleSymmetryTransformation}, requires simultaneous shift to the field. That connects the two symmetries of free scalar field theory in a non-trivial and fixed way.

Let us now see the effect of the center on the transformations of the EMT. From \eqref{eq:LiouvilleNoetherCharge} and \eqref{eq:LiouvilleChargeAlgebra} it follows that
\begin{equation}
  \pb{\tilde \Theta_{\pm\pm}(x)}{\tilde \Theta_{\pm\pm}(y)}\eval_{x^\mp=y^\mp}
  =
  \tilde\Theta'_{\pm\pm}(x)\delta(x^\pm-y^\pm)
  +2\tilde\Theta_{\pm\pm}(x)\delta'(x^\pm-y^\pm)
  -\frac2{\beta^2}\delta'''(x^\pm-y^\pm)\,,
\end{equation}
and the term proportional to $\delta'''$ stems directly from the presence of the central charge. This, through \eqref{eq:NoetherChargeAsGenerator} with \eqref{eq:LiouvilleNoetherCharge}, immediately gives the infinitesimal conformal transformation of the EMT
\begin{equation}\label{eq:InfinitesimalEMTtransformation}
  \tilde\delta_f  \tilde\Theta_{\pm\pm}
  = f^\pm\partial_\pm \tilde\Theta_{\pm\pm}
  +2 \tilde\Theta_{\pm\pm}\partial_\pm f^\pm
  -\frac2{\beta^2}\partial^3_\pm f^\pm
  \,,
\end{equation}
which is directly affected by the central charge. Thus, the central charge can be revealed by a direct computation of $\delta_f \tilde\Theta_{\mu\nu}$ already at the Lagrangian level, in analogy to \eqref{eq:LagrangianCenter} and \eqref{NewResult1}.

The infinitesimal transformation \eqref{eq:InfinitesimalEMTtransformation} can be ``exponentiated'' to a finite conformal transformation $x^\pm\goto x'^\pm = x^\pm - f^\pm(x^\pm)$ which gives the following interesting result
\begin{equation}\label{eq:EMTtransformationExponentiated}
  \tilde\Theta'_{\pm\pm}(x') = \qty(\pdv{x^\pm}{x'^\pm})^2 \qty[\tilde\Theta_{\pm\pm}(x)
  -\frac{2}{\beta^2}(\mathcal{S}x'^\pm)(x^\pm)] \,,
\end{equation}
with $(\mathcal S x')(x)$ the Schwarzian derivative
\begin{equation}\label{eq:SchwarzianDerivative}
  (\mathcal{S}g)(x)\equiv
  \frac{g'''(x)}{g'(x)}
  -\frac32\qty(\frac{g''(x)}{g'(x)})^2 \,.
\end{equation}

An interesting consequence for Liouville theory is that there exist a conformal transformation such that the EMT vanishes, as we now show. Since the solution to \eqref{eq:FlatLiouvilleEoM} can be written as \citep{Liouville_1853}
\begin{equation}
  \Phi(x) = \frac1\beta\ln\frac{F'(x^+)G'(x^-)}{\qty[1+\frac14 m^2 F(x^+)G(x^-)]^2} \,,
\end{equation}
where $F$ and $G$ are two arbitrary smooth enough functions, the EMT \eqref{eq:FlatLiouvilleEMTimproved} is
\begin{align}
  \tilde \Theta_{++} \approx &~
  3\qty(\frac{F''}{F'})^2 - 2 \frac{F'''}{F'} = -2(\mathcal S F) (x^+) \,,
  \\
  \tilde \Theta_{--} \approx &~
  3\qty(\frac{G''}{G'})^2 - 2 \frac{G'''}{G'} = -2(\mathcal S G) (x^-) \,.
\end{align}
This can clearly be canceled by a suitable conformal transformation \eqref{eq:EMTtransformationExponentiated}.


\subsection{On the general case}

We end this Section by exploring the general validity of the results just illustrated. Let us consider a two-dimensional theory in Minkowski spacetime\footnote{Analogous results hold in Euclidean spacetimes.} enjoying \textit{anomalous} conformal symmetry.

Poincaré symmetry requires the EMT, $\Theta_{\mu\nu}$, to be symmetric, $\Theta_{\mu\nu}\approx\Theta_{\nu\mu}$, and conserved, $\partial_\mu\Theta^{\mu\nu} \approx 0$, on-shell, whereas invariance under scale (and special conformal) transformations implies zero-trace on-shell, $\Theta^\mu{}_\mu\approx 0$. From these it follows that
\begin{equation}
  \partial_\mp \Theta_{\pm\pm}\approx 0\,.
\end{equation}
Therefore, not only does $f^\mu$ split, as a result of \eqref{eq:ConfKillingEqChapter2}, but the EMT as well splits into right-moving and left-moving components. This holds for \textit{any} two-dimensional conformally symmetric theory, whether or not the equations of motion imply the split of fields. Therefore, this result does not depend from the details of the theory, but only from its symmetries.

This property is immediately passed on to the Noether charges, through their very definition \eqref{eq:NoetherCharge} and
\eqref{eq:BesselHagenForm}, so they also split in general
\begin{equation}
  Q^\pm[f] \approx \int \dd{x^\pm} \Theta_{\pm\pm}(x^+)\,f^\pm(x^+) \,,
\end{equation}
and they are always independent from the ``time'' variable $x^\mp$, on-shell. Another, alternative argument is that Noether charges should be expected to split also from the fact that conformal generators split, see \eqref{lnMinkowski}, and through \eqref{eq:NoetherChargeAsGenerator} they are directly connected to the charges.

As previously shown, if a theory is anomalous, a center must take care of that in the algebra obeyed by the charges $Q[f]$, which is a centrally extended version of the algebra of generators, \eqref{eq:AlgebrasConnectionGeneral}. For conformal symmetry the generators have to follow the Witt algebra \eqref{WittMinkowski}
\begin{equation*}
  \comm{l_n^\pm}{l_m^\pm} = (n-m)\,l_{n+m}^\pm\,,
\end{equation*}
and the only possible central extension of the Witt algebra is the Virasoro algebra, \citep{MathIntroCFT_Schottenloher}. Therefore, the only possible conformally anomalous form of the algebra of charges is\footnote{This is true up to trivial central charges, that can be removed. See Section \ref{sec:ClassicalAnomalies} and also \eqref{eq:ImprovedChargeShift}.}
\begin{equation}
  i\pb{Q^\pm_n}{Q^\pm_m} = (n-m)Q^\pm_{n+m} + \frac{c}{12}n(n^2-1)\delta_{m+n,0} \,.
\end{equation}
The interdependence between charges and the EMT implies that the above propagates to the Poisson brackets of the EMT as follows\footnote{In the Euclidean the factor of $\delta'''$ is $c/12$. The extra $2\pi$ factor in the denominator, in the following results, is due to the different choices of bases, the trigonometric basis for Minkowski, \eqref{eq:MinkowskianBasis}, and the ``power'' basis for Euclidean, \eqref{eq:EuclideanBasis}. See Appendix \ref{app:EuclideMinkowski}.}
\begin{equation}
  \pb{\Theta_{\pm\pm}(x)}{\Theta_{\pm\pm}(y)}\eval_{x^\mp=y^\mp}
  =
  \Theta'_{\pm\pm}(x)\delta(x^\pm-y^\pm)
  +2\Theta_{\pm\pm}(x)\delta'(x^\pm-y^\pm)
  -\frac c{24\pi}\delta'''(x^\pm-y^\pm) \,.
\end{equation}
Using this, again it follows from \eqref{eq:NoetherChargeAsGenerator} that the infinitesimal conformal transformation of the EMT
\begin{equation}
  \delta_f\Theta_{\pm\pm}=\pb{\Theta_{\pm\pm}}{Q[f]}\,,
\end{equation}
for a generic anomalous theory necessarily involve the central charge
\begin{equation}
  \delta_f  \Theta_{\pm\pm}
  = f^\pm\partial_\pm \Theta_{\pm\pm}
  +2 \Theta_{\pm\pm}\partial_\pm f^\pm
  -\frac c{24\pi}\partial^3_\pm f^\pm \,.
\end{equation}
This is a general consequence of the fact that the symmetry algebra is the Virasoro, rather than the Witt, hence it holds for \textit{any} two-dimensional conformally anomalous theory, regardless of the details of the theory. Noticeably, this is the very same expression one obtains in the \textit{quantum} context, see, e.g., \citep{CFTDiFrancesco}. Here, though, we are fully \textit{classical}.

The ``exponentiation'' of the infinitesimal transformation to a finite conformal transformation $x^\pm\goto x'^\pm = x^\pm - f^\pm(x^\pm)$ leads to
\begin{equation}\label{eq:EMTtransformationExponentiated}
  \tilde\Theta'_{\pm\pm}(x') = \qty(\pdv{x^\pm}{x'^\pm})^2 \qty[\tilde\Theta_{\pm\pm}(x)
  -\frac{c}{24\pi}(\mathcal{S}x'^\pm)(x^\pm)] \,,
\end{equation}
with $(\mathcal S x')(x)$ the Schwarzian derivative, \eqref{eq:SchwarzianDerivative}. This result too holds for any conformally anomalous classical theory. The EMT does not transform as a tensor under a general conformal transformation due to the center, i.e., it is not a \emph{primary field} \citep{CFTDiFrancesco}. Nonetheless, it does so for M\"obius transformations
\begin{equation}
  x'^\pm = \frac{ax^\pm + b}{dx^\pm + e} \,,
\end{equation}
where $ae-bd \neq 0$, because these transformations have vanishing Schwarzian derivative and are generated through charges $Q^\pm_{-1}, Q^\pm_0,  Q^\pm_{1}$ which close a centerless subalgebra isomorphic to the global conformal algebra \citep{CFTDiFrancesco}.


\section{Liouville theory on curved backgrounds} \label{Sec:LiouvilleCurved}

Knowing that, in the quantum case, anomalous EMTs in flat space are responsible for interesting phenomena when the theory is embedded in curved backgrounds, it is time we
embed classical Liouville theory in a curved spacetime. Although this has been done extensively in the literature, only recently we have shown \cite{Letter1HamanIorio2023} that in this process a gravitational anomaly is hiding, as suspected by Jackiw \citep{WeylvsLiouville}. In particular, our goal here is to explicitly show the details of the behavior of the EMT under diffeomorphism and Weyl transformations.

Throughout the Section, several versions of Liouville action in curved spacetime are scrutinized, and we list them all in Appendix \ref{app:Liouvilles}. Let us start with the diffeomorphism invariant version of flat\footnote{
  We should use different indices for flat spacetime, where form invariance holds under Lorentz transformations, say Latin letters $a, b, c, \ldots$, and curved spacetime, where form invariance holds under general coordinate transformations, say Greek letters $\mu,\nu,\lambda,\ldots$. Here we do not do that since all four flat-related situations are specifically addressed and no ambiguity should arise. Details on how to deal with different indices are given in the Appendix \ref{app:Vielbeins}.}
  Liouville action \eqref{eq:FlatLiouvilleTheory}
\begin{equation}\label{eq:MinimalLiouville}
  \mathcal A_{\scriptscriptstyle{D}}[\Phi] = \int \dd[2]{x}\sqrt{-g}
  \qty(
    \frac{1}{2}g^{\mu\nu}\nabla_\mu\Phi\nabla_\nu\Phi
    - \frac{m^2}{\beta^2}e^{\beta\Phi}) \,,
\end{equation}
with equation of motion
\begin{equation}
  g^{\mu\nu}\nabla_\mu\nabla_\nu\Phi + \frac{m^2}{\beta}e^{\beta\Phi} = 0 \,,
\end{equation}
and EMT, obtained as in \eqref{eq:CurvedEMTensor}, given by
\begin{equation}
  T^{\mu\nu}_{\scriptscriptstyle{D}} = \nabla^\mu\Phi\nabla^\nu\Phi -g^{\mu\nu}
  \qty(
    \frac12g^{\alpha\beta}\nabla_\alpha\Phi\nabla_\beta\Phi
    - \frac{m^2}{\beta^2}e^{\beta\Phi}
  ) \,.
\end{equation}
This tensor is covariantly conserved, $\nabla_\mu T^{\mu\nu}_{\scriptscriptstyle{D}} \approx 0$, and symmetric, but its trace is non-vanishing
\begin{equation}
  {T_{\scriptscriptstyle{D}}}^\mu{}_\mu = \frac{2m^2}{\beta^2} e^{\beta\Phi} \,,
\end{equation}
even in the flat limit, which clashes with the just discussed conformal invariance of previous Section.

The simplest \emph{non-minimal, ad hoc} improvement of the action is
\begin{equation}\label{eq:CurvedLiouville}
  \mathcal A{}_{\scriptscriptstyle{NM}}[\Phi]=\int\dd[2]{x}\sqrt{-g}
  \qty(
    \frac12g^{\mu\nu}\nabla_\mu\Phi\nabla_\nu\Phi
    - \frac{m^2}{\beta^2}e^{\beta\Phi}
    + \alpha R\Phi
  ) \,,
\end{equation}
where $\alpha$ is a constant real parameter and $R = g^{\sigma\nu}R^\rho{}_{\sigma\rho\nu}$ is the Ricci scalar, descending from the Riemann tensor
${R^\rho}_{\sigma\mu\nu} \equiv
  \partial_\mu {\Gamma^\rho}_{\nu\sigma}
  - \partial_\nu {\Gamma^\rho}_{\mu\sigma}
  + {\Gamma^\rho}_{\mu\lambda}{\Gamma^\lambda}_{\nu\sigma}
  - {\Gamma^\rho}_{\nu\lambda}{\Gamma^\lambda}_{\mu\sigma}$. The equation of motion is
\begin{equation}\label{eq:CurvedLiouvilleEquation}
  g^{\mu\nu}\nabla_\mu\nabla_\nu\Phi + \frac{m^2}{\beta}e^{\beta\Phi}
  =   \alpha R \,,
\end{equation}
and the EMT is ``improved'' to
\begin{equation}\label{eq:CurvedEMTLiouville}
    T^{\mu\nu}_{\scriptscriptstyle{NM}} =~ \nabla^\mu\Phi\nabla^\nu\Phi -g^{\mu\nu}
    \qty(\frac12g^{\alpha\beta}\nabla_\alpha\Phi\nabla_\beta\Phi
      - \frac{m^2}{\beta^2}e^{\beta\Phi})
    + 2\alpha\qty(g^{\mu\nu} \nabla_\rho\nabla^\rho - \nabla^\mu\nabla^\nu)\Phi \,.
\end{equation}
Comparing its flat limit with the flat EMT \eqref{eq:FlatLiouvilleEMTimproved} gives the additional condition $\alpha = 1 / \beta$. The trace is still non-vanishing for a general metric
\begin{equation}\label{eq:CurvedEMTtrace}
  {T_{\scriptscriptstyle{NM}}}^\mu{}_\mu \approx \frac2{\beta^2} R \,,
\end{equation}
however it does vanish in the flat limit, so \eqref{eq:CurvedEMTLiouville}, with $\alpha=1/\beta$ is the correct EMT\footnote{The subscript ``$L$'' refers to the commonly used Liouville theory, discussed later and in the Appendix \ref{app:Liouvilles}.} $T_{\scriptscriptstyle{L}}^{\mu\nu}$
\begin{equation}\label{eq:CurvedEMTLiouvilleCorrect}
    T^{\mu\nu}_{\scriptscriptstyle{L}} =~ \nabla^\mu\Phi\nabla^\nu\Phi -g^{\mu\nu}
    \qty(
      \frac12g^{\alpha\beta}\nabla_\alpha\Phi\nabla_\beta\Phi
      - \frac{m^2}{\beta^2}e^{\beta\Phi} )
      + \frac2\beta\qty(g^{\mu\nu} \nabla_\rho\nabla^\rho - \nabla^\mu\nabla^\nu)\Phi \,.
\end{equation}

Thus flat Liouville \eqref{eq:FlatLiouvilleTheory} enjoys Poincaré symmetry as well as scale and conformal symmetries, while the curved Liouville in \eqref{eq:CurvedLiouville} is diffeomorphism invariant, but the trace \eqref{eq:CurvedEMTtrace} shows that to promote scale and conformal symmetry to curved spacetime requires additional efforts.

All of the above is classical, but still related to the quantum trace anomalies in two dimensions \cite{QFTCurvedSpace}
\begin{equation}\label{eq:AnomalousTrace}
  T^\mu{}_\mu = \frac{c}{24\pi} R \,,
\end{equation}
since the known form of the flat EMT \eqref{eq:FlatLiouvilleEMTimproved} requires the additional $R\Phi$ term in the action to obtain the improvement
$\frac{2}{\beta}\qty(
      g^{\mu\nu} \nabla_\rho\nabla^\rho - \nabla^\mu\nabla^\nu
    )\Phi$,
which survives the flat limit, although the $R\Phi$ term itself does not. It is this improvement which reflects the presence of central charge
\begin{equation}
  c = \frac{48\pi}{\beta^2},
\end{equation}
of the Virasoro algebra. Thus, the non-minimal coupling to the curvature $R\Phi$ introduces the central charge in the action and the coupling constant $\alpha$ has to be proportional to the center $c$. At the same time, the term $\alpha R\Phi$ modifies the trace of \eqref{eq:CurvedEMTensor} to the exact form \eqref{eq:AnomalousTrace}.
It can be proved that, for any two-dimensional conformal quantum field theory, the existence of Virasoro algebra in the flat spacetime implies the trace of the EMT to be as of \eqref{eq:AnomalousTrace}, \citep{StringTheoryNutshell_Kiritsis}.

\subsection*{Weyl gauging}

The natural generalization of the scale transformation to curvilinear coordinates is the rigid Weyl transformation \eqref{eq:WeylTransformationMetric}. For Liouville this gives
\begin{equation}\label{eq:RicciWeylGauging2D}
    g_{\mu\nu} \weyl e^{2\omega}g_{\mu\nu} \,, \qquad \Phi \weyl \Phi-\frac2\beta\omega \,.
\end{equation}
Notice that, while the Weyl transformation of the metric is always the same, no matter the spacetime dimensions or the theory, the specific form of the Lagrangian can change dramatically the Weyl transformation of the fields. In this case, it is the exponential term in \eqref{eq:MinimalLiouville} (the potential) that dictates how $\Phi$ has to transform in \eqref{eq:RicciWeylGauging2D}. Indeed, the \emph{field independent} shift there is what makes $\sqrt{-g}\exp\qty(\beta\Phi)$ a Weyl scalar\footnote{This is not the case for the free scalar theory in \eqref{eq:FreeScalarTheory}. If we write in curvilinear coordinates the only (kinetic) term there, $\sqrt{-g} g^{\mu\nu}\partial_\mu\varphi\partial_\nu\varphi$, we notice that in two dimensions $\sqrt{-g} g^{\mu\nu}$ is a Weyl scalar, so the field $\varphi$ can be taken as a Weyl scalar, hence not transforming at all. On the other hand, in analogy with the logic that brought us to \eqref{eq:FreeImprovedConformalTransformation}, since only derivatives of $\varphi$ appear, one is free to define the Weyl transformation of $\varphi$ as a constant field-independent shift. Again, while for the free scalar this shift is not necessary to have rigid Weyl symmetry, for Liouville it is.}.

It can be directly verified that both actions, \eqref{eq:MinimalLiouville} and \eqref{eq:CurvedLiouville}, are invariant under these rigid transformations.
Promoting $\omega$ to $\omega(x)$, is the first step of the \textit{Weyl-gauging} procedure, an ancestor of modern gauging \citep{GaugeTheory}.

The combination of a diffeomorphism leaving the metric rescaled
\begin{equation}
  g'_{\mu\nu}(x') =
  \pdv{x^\alpha}{x'^\mu}\pdv{x^\beta}{x'^\nu} g_{\alpha\beta}(x) = \Omega\qty(x(x'))g_{\mu\nu}\qty(x(x')) \,,
\end{equation}
with a Weyl transformation that gets rid of the factor $\Omega$, leaves the metric untouched and it might be a symmetry. Such transformation coincides with the definition of conformal transformation \eqref{eq:ConformalKilling} in curved spacetime. Then, the simultaneous invariance under Weyl transformation and diffeomorphism implies the conformal symmetry. In the case of Liouville theory, the conformal symmetry transformation \eqref{eq:LiouvilleSymmetryTransformation} is
\begin{equation}
  \tilde\delta_f \Phi = \underbrace{\vphantom{\frac1\beta}f^\mu\partial_\mu\Phi}_{\text{Diff}} + \underbrace{\frac1\beta\partial_\mu f^\mu}_{\text{Weyl}} \,,
\end{equation}
with the Weyl transformation parameter in \eqref{eq:RicciWeylGauging2D} restricted by the conformal diffeomorphism to be
\begin{equation}
  \omega = -\frac{1}{2}\partial_\mu f^\mu \,.
\end{equation}

In the process of gauging, a Weyl covariant derivative $D_\mu$ has to be introduced. The choice
\begin{equation}
  \nabla_\mu\Phi\goto D_\mu\Phi \equiv \nabla_\mu\Phi - \frac{2}{\beta}W_\mu \,,
\end{equation}
where $W_\mu$ is the (diffeomorphism covariant) Weyl gauge vector field, transforming under Weyl transformation as
\begin{equation}\label{eq:WeylGFWeylTransformation}
  W_\mu\weyl W_\mu - \partial_\mu\omega \,,
\end{equation}
ensures $D_\mu\Phi$ to be a Weyl scalar. The field-independent term in the Weyl covariant derivative, $D_\mu$, is the direct consequence of the field-independent shift in Liouville field \eqref{eq:RicciWeylGauging2D}, discussed there. This promotes curvilinear Liouville theory to the \textit{Weyl-gauged} action
\begin{equation}\label{eq:WeylGaugedLiouville}
  \mathcal A_{\scriptscriptstyle{W}}[\Phi,W_\mu]
  = \int \dd[2]{x}\sqrt{-g}
  \qty(
    \frac{1}{2}g^{\mu\nu}\nabla_\mu\Phi\nabla_\nu\Phi
    - \frac{m^2}{\beta^2}e^{\beta\Phi}
    +\frac{2}{\beta}\Phi\nabla_\mu W^\mu
    +\frac{2}{\beta^2}g^{\mu\nu}W_\mu W_\nu
    ) \,.
\end{equation}

The action $\mathcal A_{\scriptscriptstyle{W}}$ is invariant under both diffeomorphism and Weyl transformations, at the price of the appearance of the Weyl field. This is undesirable as the goal is to formulate Liouville theory solely in terms of the dynamical $\Phi$ and geometric tensors. In \citep{WeylGauging} it is shown that, for flat scale invariant theories that also enjoy full conformal symmetry, the expressions in terms of Weyl fields can be traded of for expressions in terms of curvature tensors, a process called there \textit{Ricci-gauging}.

Indeed certain combinations of the Weyl field and its derivatives transform like curvature  tensors. In two dimensions
\begin{equation}\label{eq:2DWeylGaugingCombo}
  2\nabla_\mu W^\mu \weyl e^{-2\omega}\qty(2\nabla_\mu W^\mu - 2\nabla_\mu\nabla^\mu \omega) \,,
\end{equation}
should be compared to
\begin{equation}\label{eq:2DRicciGaugingCombo}
  R[g_{\mu\nu}]\weyl e^{-2\omega}\qty(R[g_{\mu\nu}]-2\nabla_\mu\nabla^\mu \omega) \,,
\end{equation}
while in higher dimensions, $d>2$,
\begin{equation}\label{eq:WeylGaugingCombo}
  \Omega_{\mu\nu}[W]\equiv \nabla_\mu W_\nu-W_\mu W_\nu +\frac12 g_{\mu\nu}W_\rho W^\rho \,,
\end{equation}
transforms identically to
\begin{equation}\label{eq:RicciGaugingCombo}
  \frac{1}{d-2}\qty[R_{\mu\nu}-\frac{1}{2(d-1)}g_{\mu\nu}R] \,,
\end{equation}
that is
\begin{equation}
  \Omega_{\mu\nu}[W]
  \weyl
  \Omega_{\mu\nu}[W]
  -\nabla_\mu\nabla_\nu\omega+\partial_\mu\omega\partial_\nu\omega-\frac12g_{\mu\nu}\partial_\rho\omega\partial^\rho\omega \,.
\end{equation}
Thus, if $W_\mu$ appears in the Weyl-gauged theory \emph{only} in the previous combinations \eqref{eq:2DWeylGaugingCombo} or \eqref{eq:WeylGaugingCombo}, their trading for \eqref{eq:2DRicciGaugingCombo} or \eqref{eq:RicciGaugingCombo}, respectively, does not violate the invariance under diffeomorphism and Weyl transformations. Yet, it does change the equations of motion, hence, the meaning of the theory. Furthermore, this allows to systematically generate improvements for the EMT, then computed via \eqref{eq:CurvedEMTensor}.

An implicit argument in \citep{WeylGauging} suggests that the trading of \eqref{eq:2DWeylGaugingCombo} for \eqref{eq:2DRicciGaugingCombo} always provides a theory invariant under rigid Weyl and local Weyl transformations, up to field independent terms. If we drop such terms, though, we lose track of the central charge $c$, that is an important task here. We cannot omit terms that may actually be the manifestations of non-trivial centers. They actually are the focus of the rest of this paper.

The trading of $2\nabla_\mu W^\mu$ for $R$ introduces the non-minimal coupling, with correct factor $1/\beta$, in the Weyl-gauged action \eqref{eq:WeylGaugedLiouville}
\begin{equation}\label{eq:RicciGaugedLiouville}
    \mathcal A_{\scriptscriptstyle{R}}[\Phi] = \int \dd[2]{x}\sqrt{-g}
    \qty(
      \frac{1}{2}g^{\mu\nu}\nabla_\mu\Phi\nabla_\nu\Phi
      - \frac{m^2}{\beta^2}e^{\beta\Phi}
      +\frac{1}{\beta}\Phi R
      +\frac{2}{\beta^2}g^{\mu\nu}W_\mu W_\nu) \,,
\end{equation}
but here, contrary to what is done in \citep{WeylGauging}, we keep the term independent from $\Phi$ which ensures the Weyl invariance, provided
\begin{equation}\label{eq:WeylRicci2DIdentification}
  2\nabla_\mu W^\mu = R \,,
\end{equation}
holds as an equation for $W^\mu$. Moreover, the equation of motion for $\Phi$ now reproduces the standard one in \eqref{eq:CurvedLiouvilleEquation} with the coupling $\alpha=1/\beta$
\begin{equation}
  g^{\mu\nu}\nabla_\mu\nabla_\nu\Phi + \frac{m^2}{\beta}e^{\beta\Phi}
  =
  \frac1\beta R \,,
\end{equation}
and we also recover the right flat limit of \eqref{eq:FlatLiouvilleEoM}, with no additional demand of vanishing Weyl field.

Since anomalies are routinely considered as purely quantum phenomena, hence the concerns of this paper are not shared by many, it is customary to simply drop $W^\mu W_\mu$. In fact, it does not affect the equation of motion and the quantum trace anomaly is to be expected. That is why the most common form of Liouville action in curved spacetime is \citep{WeylGauging}
\begin{equation} \label{eq:CommonLiouville}
  \mathcal A_{\scriptscriptstyle{L}}[\Phi]=\int\dd[2]{x}\sqrt{-g}
  \qty(
    \frac12g^{\mu\nu}\nabla_\mu\Phi\nabla_\nu\Phi
    -\frac{m^2}{\beta^2}e^{\beta\Phi}
    +\frac1\beta R\Phi
  ) \,,
\end{equation}
that is \eqref{eq:CurvedLiouville} for $\alpha=1/\beta$.
It enjoys diffeomorphism invariance, and, in the flat spacetime, conformal symmetry. It also gives rise to the EMT $T_{\scriptscriptstyle{L}}^{\mu\nu}$ in \eqref{eq:CurvedEMTLiouvilleCorrect}.

Nonetheless, the question of what to make of this extra term is of great importance in this paper and the subject matter of the next Section.

\section{The extra term $W^\mu W_\mu$} \label{Sec:TheExtraWTerm}

Weyl gauging provided yet another additional term, proportional to $W^\mu W_\mu$, to the curvilinear Liouville action. The independence from the field $\Phi$ and the request $2\nabla_\mu W^\mu = R$, reduce this extra term to a purely geometrical term, which cannot affect the equation of motion \eqref{eq:CurvedLiouvilleEquation}. This goes hand in hand with the fact that, under Weyl transformation of the action $\mathcal A_{\scriptscriptstyle{R}}$ in \eqref{eq:RicciGaugedLiouville}, this term provides the cancellation of the extra field-independent contributions.

The restored Weyl invariance means that the trace of the EMT vanishes again. Since the $W^\mu$ is the sole cause of that it is clear that the $W^\mu W_\mu$ term carries the center in the form of the ``mass''\footnote{
  In the gauged version of the theory the $W^\mu W_\mu$ term gives the mass to the field $W^\mu$, as the term is accompanied by mass factor $m^2/2$.
} as it has to cancel the trace \eqref{eq:AnomalousTrace} also proportional to the center. In fact, comparing with \eqref{eq:CentralChargeLiouville} the extra term is exactly
\begin{equation}\label{eq:ExtraTerm}
  \frac{c}{24\pi}W^\mu W_\mu \,.
\end{equation}

This implies that either the trace anomaly is a purely quantum effect or that this extra term cancels the center completely from the theory. The impact on the Virasoro algebra in flat spacetime can be understood with no computations. The term \eqref{eq:ExtraTerm} provides only field independent improvement to the EMT. If it survives in the flat limit (similarly to the improvement generated by $R\Phi$ term) this improvement is invisible to the Poisson brackets which implies that this additional improvement gives rise only to \emph{trivial} central charge and does not change the \emph{genuine} center $c$ of the Virasoro algebra in \eqref{eq:CentralChargeLiouville}.

To understand how \eqref{eq:ExtraTerm} ensures the tracelessness and to see the subtleties accompanying this hunt for Weyl invariance one has to look under the hood of \eqref{eq:WeylRicci2DIdentification} and solve the equation.

A solution of \eqref{eq:WeylRicci2DIdentification} can be provided in the language of Green's functions \citep{WeylvsLiouville}, i.e. by finding $K(x,y)$ such that
$\nabla^2_x K(x,y) = \frac{1}{\sqrt{-g(x)}}\delta^{(2)}(x-y)$. Assuming that $W_\mu$ is a gradient of a scalar field $w$, transforming as $w\weyl w -\omega$, the solution to the \eqref{eq:WeylRicci2DIdentification} is
\begin{equation}
  w(x) = \frac12 \int\dd[2]{y}\sqrt{-g(y)}K(x,y)R(y) \,.
\end{equation}
It follows that the extra $W^\mu W_\mu$ term in the action is
\begin{equation}\label{eq:PolyakovEffectiveAction}
  \int \dd[2]{x} W^\mu(x) W_\mu(x)
  =
  \frac14\int \dd[2]{x}\dd[2]{y} \sqrt{-g(x)}R(x)K(x,y)\sqrt{-g(y)}R(y) \,,
\end{equation}
which is the well-known \textit{Polyakov effective action} \citep{Polyakov_1981}, describing a quantized two-dimensional bosonic string with vanishing cosmological constant.

This action appears in various places in physics, namely in the context of two-dimensional gravity, where one of the first proposed models was the Liouville gravity, \citep{QuantumGravity_Christensen}, governed by \eqref{eq:2Dgravity}
\begin{equation}\label{eq:LiouvilleGravity}
  R - \Lambda = 0 \,,
\end{equation}
where $\Lambda$ is the cosmological constant. Polyakov effective action yields this equation provided an additional cosmological term, $\int\dd[2]{x} \sqrt{-g}\Lambda$ is added.

Correspondingly, the Polyakov effective action provides a simple model exhibiting the trace anomaly, since it is diffeomorphism but not Weyl invariant\footnote{This is to be expected as it was derived here from the $W^\mu W_\mu$ improvement of exactly such transformation properties.}\footnote{In the quantum case, there is no regularization for which Weyl and diffeomorphism invariance hold together \citep{LEUTWYLER}.}, \citep{Torre_1989, Karakhanyan_1994, Jackiw_1995}. Moreover, the trace of the EMT derived from the Polyakov effective action \eqref{eq:PolyakovEffectiveAction} has the same form as the trace of Liouville EMT \eqref{eq:CurvedEMTtrace}.

Such solution, then gives
\begin{equation} \label{eq:PolyakovLiouville}
  \begin{split}
    \mathcal A_{\scriptscriptstyle{LP}}[\Phi] =&~\int\dd[2]{x}\sqrt{-g}
    \qty(
      \frac12g^{\mu\nu}\nabla_\mu\Phi\nabla_\nu\Phi
      -\frac{m^2}{\beta^2}e^{\beta\Phi}
      +\frac1\beta R\Phi
    )
    \\
    &+\frac1{2\beta^2} \int \dd[2]{x}\dd[2]{y} \sqrt{-g(x)}R(x)K(x,y)\sqrt{-g(y)}R(y)\,,
  \end{split}
\end{equation}
which is both diffeomorphism and Weyl invariant, at the price of the \emph{loss of locality}.

From now on we shall no longer consider the Polyakov effective action, as our focus is on \textit{local} solutions to \eqref{eq:WeylRicci2DIdentification}. In particular, we shall delve into the local solution that was found by Deser and Jackiw in \citep{DeserJackiw96}. In fact, these authors arrived at their local solution by looking into the Polyakov action and its associated EMT\footnote{A particular solution for Reissner-Nordström black hole was derived by Iso et al in \citep{Iso2006PRD} using the Green function approach.}. The solution in point is
\begin{equation}\label{eq:JackiwSolution}
    W^\mu_{\scriptscriptstyle{DJ}} =
    \frac{\varepsilon^{\mu\nu}}{2\sqrt{-g}}\qty[
    \frac{\varepsilon^{\alpha\beta}}{\sqrt{-g}}
    \Gamma_{\beta\alpha\nu} 
    +
    (\cosh\sigma -1)\partial_\nu\gamma ]\,,
\end{equation}
where $\varepsilon^{tx}=+1$ is the Levi-Civita symbol\footnote{Alternatively, we choose $\varepsilon^{-+}=+1$ in the light-cone coordinates as explained in Appendix \ref{app:EuclideMinkowski}.} and a ``conformal'' parametrization of the metric is introduced
\begin{equation}\label{eq:MetricParametrization}
  \frac{g_{++}}{\sqrt{-g}} = e^\gamma\sinh\sigma \,, \qquad
  \frac{g_{+-}}{\sqrt{-g}} = \cosh\sigma \,, \qquad
  \frac{g_{--}}{\sqrt{-g}} = e^{-\gamma}\sinh\sigma \,,
\end{equation}
and from there
\begin{equation}\label{eq:GammaParametrization}
  \gamma = \ln\sqrt{g_{++}/{g_{--}}} \,.
\end{equation}

Two peculiar properties might be immediately spotted. First, the equation \eqref{eq:WeylRicci2DIdentification} is blind to additional improvements of the form
\begin{equation}\label{eq:WAmbiguity}
  W^\mu_{\scriptscriptstyle{DJ}} \goto W^\mu_{\scriptscriptstyle{DJ}} + \frac{\varepsilon^{\mu\nu}}{2\sqrt{-g}}\partial_\nu r,
\end{equation}
where $r$ is any Weyl scalar. Thus, the solution \eqref{eq:JackiwSolution} is not unique. Second, the first term of $W_{\scriptscriptstyle{DJ}}^\mu$ is proportional to the Levi-Civita connection while the second term requires parametrization of the metric. This indicates non-vectorial transformation properties. In fact, under an infinitesimal diffeomorphism
\begin{equation}\label{eq:InfinitesimalDiffeo}
  x^\mu\goto x^\mu - f^\mu(x),
\end{equation}
it transforms as
\begin{equation}\label{eq:JackiwSolutionDiffeoTransf}
    W'{}_{\scriptscriptstyle{DJ}}^\mu(x') = ~\pdv{x'^\mu}{x^\nu}W_{\scriptscriptstyle{DJ}}^\nu(x)
    + \frac{\varepsilon^{\mu\nu}}{2\sqrt{-g}}\partial_\nu
    \qty[
      \qty(
        \partial_--e^{-\gamma}\tanh \frac\sigma2\partial_+
      )f^-
      -
      \qty(
        \partial_+-e^{\gamma}\tanh \frac\sigma2\partial_-
      )f^+
    ] \,,
\end{equation}
as shown in  \citep{Haman_2019}. This proves that the local solution to $2\nabla_\mu W^\mu =R$ is not a contravariant vector, even though its divergence $\nabla_\mu W^\mu_{\scriptscriptstyle{DJ}}$ is still a scalar. On the other hand, under Weyl transformation
\begin{equation}
  g_{\mu\nu}W_{\scriptscriptstyle{DJ}}^\nu \weyl g_{\mu\nu}W_{\scriptscriptstyle{DJ}}^\nu - \partial_\mu\omega \,.
\end{equation}
That is the required transformation \eqref{eq:WeylGFWeylTransformation}.

A solution that bypasses the parametrization \eqref{eq:MetricParametrization} is possible. It requires the introduction of \emph{vielbein} and of \emph{spin connection} (see Appendix \ref{app:Vielbeins}). If one does that $W^\mu$ takes a compact and appealing form
\begin{equation}
  W^\mu =
  \frac{\varepsilon^{\mu\nu}}{2\sqrt{-g}}
  s_{\nu ab}\varepsilon^{ab} \,,
\end{equation}
where $s_{\nu ab}$ is the spin connection, the Latin indices $a,b$ represent the local Lorentz frame and the Greek indices $\mu,\nu$ represent the general coordinate frame as is explained in detail in Appendix \ref{app:Vielbeins}. Again it can be seen that $W^\mu$ is not a contravariant vector.

It follows from \eqref{eq:JackiwSolutionDiffeoTransf} that the improvement term $g_{\mu\nu}W_{\scriptscriptstyle{DJ}}^\mu W_{\scriptscriptstyle{DJ}}^\nu$ of the action cannot be a scalar anymore and while this solution keeps Weyl invariance and locality of the whole action \eqref{eq:RicciGaugedLiouville}, it breaks the diffeomorphism invariance.

~

The handling of the $W^\mu W_\mu$ term closely resembles the situation one encounters dealing with \textit{quantum} trace anomaly, even though we never left the \textit{classical} theory. Here and there, one is forced to keep only one of the two invariances, Weyl or diffeomorphism, but cannot have both, \citep{Guadagnini_1988}. In the rest of the paper the implications and the ``amount'' of diffeomorphism breakdown are studied, focusing on the $W_{\scriptscriptstyle{DJ}}^\mu$ solution
\begin{equation}\label{eq:GeneralestW}
  W^\mu_{\scriptscriptstyle{DJ}} =
  \frac{\varepsilon^{\mu\nu}}{2\sqrt{-g}}
  \qty[
    \frac{\varepsilon^{\alpha\beta}}{\sqrt{-g}}
    \Gamma_{\beta\alpha\nu}
    +
    (\cosh\sigma -1)\partial_\nu\gamma
    +
    \partial_\nu r
  ] \,,
\end{equation}
where the term \eqref{eq:WAmbiguity}, that makes explicit the mentioned ambiguity, is included. The label ``$DJ$'' is omitted in the following.

\section{Energy\hyp{}momentum tensor} \label{Sec:EMTensor}

Let us now look at the impact of \eqref{eq:GeneralestW} on the EMT. First consider the variational identity
\begin{equation}
  \delta\qty[(\cosh\sigma-1)\partial_\nu\gamma]
  - \partial_\nu\qty[(\cosh\sigma-1)\delta\gamma]
  =
  \delta\qty(\frac{\varepsilon^{\alpha\beta}}{\sqrt{-g}})
  \Gamma_{\beta\alpha\nu}
  +
  \frac{\varepsilon^{\alpha\beta}}{\sqrt{-g}}
  \Gamma_{\beta\lambda\nu} g^{\lambda\sigma}\delta g_{\alpha\sigma}\,,
\end{equation}
giving
\begin{equation}
    \delta W^\mu =
    -\frac12 W^\mu g^{\alpha\beta}\delta g_{\alpha\beta}
    +\frac{\varepsilon^{\mu\nu}}{2\sqrt{-g}}
    \qty(
      \frac{\varepsilon^{\alpha\lambda}}{\sqrt{-g}}
      \qty(
        \Gamma^\beta{}_{\lambda\nu} \delta g_{\alpha\beta}
        +\partial_\alpha \delta g_{\nu\lambda}
      )
      +\partial_\nu \qty[\qty(\cosh\sigma -1)\delta\gamma + \delta r]
    )\,.
\end{equation}
Then the extra contribution
\begin{equation*}
  T^{\mu\nu}_{\text{extra}}\equiv
  -\frac{2}{\sqrt{-g}}\fdv{g_{\mu\nu}} \int \dd[2]{x}\frac{2}{\beta^2}\sqrt{-g}W^\mu W_\mu,
\end{equation*}
was derived in \citep{Haman_2019} as
\begin{equation}\label{eq:EMTExtraTerm}
  \begin{split}
    \beta^2 T^{\mu\nu}_{\text{extra}}
    =&~
    2g^{\mu\nu}W_\rho W^\rho
    - 4 W^\mu W^\nu
    - 2Rg^{\mu\nu}
    + 2\nabla^\mu W^\nu
    + 2\nabla^\nu W^\mu
    \\&
    + 4\frac{\varepsilon^{\alpha\beta}}{\sqrt{-g}}
    \partial_\beta W_\alpha
    [(\cosh\sigma-1)\Gamma^{\mu\nu}
    + r^{\mu\nu} 
    ]\,,
  \end{split}
\end{equation}
where $W_\rho\equiv g_{\rho\lambda}W^\lambda$ and
\begin{equation}
  \Gamma^{\mu\nu}
  =
  \frac{1}{2\sqrt{g_{--}g_{++}}}
  \begin{pmatrix}
    -\sinh \gamma & \cosh\gamma \\
    \cosh\gamma & -\sinh\gamma
  \end{pmatrix} \,,
\end{equation}
and $r^{\mu\nu}\equiv \delta r / \delta g_{\mu\nu}$.

As wanted, the trace of this improvement\footnote{Assuming $r^{\mu\nu}g_{\mu\nu} = 0$.}
\begin{equation}
g_{\mu\nu} T^{\mu\nu}_{\text{extra}} = -\frac{2}{\beta^2} R \,,
\end{equation}
exactly cancels the trace \eqref{eq:AnomalousTrace} and the complete EMT, $T_{\scriptscriptstyle{L}}^{\mu\nu} + T^{\mu\nu}_{\text{extra}}$, is traceless, proving the Weyl invariance of the theory.

The diffeomorphism invariance, that manifested itself as a conservation equation $\nabla_\mu T^{\mu\nu} = 0$, is unclear as\footnote{The covariant derivative here is computed as if the $W^\mu$ and consequently $T^{\mu\nu}_{\text{extra}}$ were proper tensors.}
\begin{equation}\label{eq:EMTDiffeoBroken}
  \begin{split}
    \beta^2\nabla_\mu T^{\mu\nu}_{\text{extra}}
    =&~
    2\nabla_\rho\nabla^\rho W^\nu
    - RW^\nu - \nabla^\nu R
    +
    4\frac{\varepsilon^{\alpha\beta}}{\sqrt{-g}}\partial_\mu
    \bigg(
      \partial_\beta W_\alpha
      \qty[
        (\cosh\sigma - 1)\Gamma^{\mu\nu}
        + r^{\mu\nu}
      ]
    \bigg)
    \\
    &+ 4\frac{\varepsilon^{\alpha\beta}}{\sqrt{-g}}\partial_\beta W_\alpha\,
    g^{\mu\nu}\partial_\sigma g_{\lambda\mu}
    \qty(
      \qty[
        (\cosh\sigma - 1)\Gamma^{\sigma\lambda}
        + r^{\sigma\lambda}
      ]
      + \frac12\frac{\varepsilon^{\sigma\lambda}}{\sqrt{-g}}
    )\,,
  \end{split}
\end{equation}
where $2\nabla_\rho\nabla^\rho W^\nu
    = 2g^{\alpha\beta}
    (\partial_\alpha\partial_\beta W^\nu
      + 2\Gamma^\nu{}_{\beta\lambda}\partial_\alpha W^\lambda
      - \Gamma^\lambda{}_{\alpha\beta}\partial_\lambda W^\nu
      +W^\lambda\partial_\lambda\Gamma^\nu{}_{\alpha\beta})
    -RW^\nu$.

Due to the parametrization \eqref{eq:MetricParametrization}, we were not able to derive a simpler form of \eqref{eq:EMTDiffeoBroken} to put in contact with known expressions of the \textit{quantum} gravitational anomalies, such as
the so-called \textit{consistent} anomaly \cite{BetlmannBOOK} (our notation follows \citep{Wilczek_2005}, see also \citep{BanerjeeKulkarni})
\begin{equation}\label{eq:WilczekDiffeoAnomaly}
  \nabla_\mu T^\mu{}_\nu = \frac{1}{48\pi}\frac{\varepsilon^{\sigma\rho}}{2\sqrt{-g}}\partial_\rho\partial_\lambda\Gamma^\lambda{}_{\nu\sigma}\,,
\end{equation}
or the so-called \textit{covariant} anomaly \cite{BetlmannBOOK} (our notation follows \citep{Jackiw_1995})
\begin{equation}\label{eq:JackiwDiffeoAnomaly}
  \nabla_\mu T^\mu{}_\nu = \frac{1}{48\pi}\partial_\nu R \,.
\end{equation}

Let us then rewrite \eqref{eq:EMTDiffeoBroken} in isothermal coordinates and let us denote all quantities evaluated in these coordinates as endowed with a hat. We have
\begin{equation}
  \hat g_{\mu\nu}(x) = e^{2\rho(x)}\eta_{\mu\nu}
  = e^{2\rho(x)}
  \begin{pmatrix}
    1 & 0 \\
    0 & -1
  \end{pmatrix} \,,
\end{equation}
where the indices are labeled $\mu=(t,r)$, and in light-cone coordinates
\begin{equation}
  \hat g_{\pm\pm}(x) = e^{2\rho(x)}
  \begin{pmatrix}
    0 & 1 \\
    1 & 0
  \end{pmatrix} \,.
\end{equation}
The Levi-Civita connection, contracted in the covariant indices, vanishes
\begin{equation}
  \hat g^{\mu\nu}\hat\Gamma{}^\lambda{}_{\mu\nu} = 0\,,
\end{equation}
as they are
\begin{equation}
  \hat\Gamma{}^{t}{}_{\mu\nu}
  =
  \begin{pmatrix}
    \dot\rho & \rho' \\
    \rho' & \dot \rho
  \end{pmatrix}
  ,\qquad
  \hat\Gamma{}^{r}{}_{\mu\nu}
  =
  \begin{pmatrix}
    \rho' & \dot\rho \\
    \dot\rho & \rho'
  \end{pmatrix},
\end{equation}
or even simpler in the light-cone version
\begin{equation}
  \hat\Gamma{}^\pm{}_{\pm\pm} = 2\partial_\pm \rho\,,
\end{equation}
where the other combinations vanish. The Ricci scalar is
\begin{equation}\label{eq:RicciScalarIsothermalCoordinates}
  \hat R = -2 \hat g^{\mu\nu}\partial_\mu\partial_\nu\rho\,.
\end{equation}
Moreover, the relation between Levi-Civita connection and Ricci scalar simplifies to
\begin{equation}
  2\hat g^{\alpha\beta}\partial_\alpha\hat \Gamma {}^\lambda{}_{\beta\mu}
  = -\hat R \hat g^\lambda{}_\mu\,,
\end{equation}
since
\begin{equation}
  \hat g{}^{\alpha\beta}\hat\Gamma{}^\lambda{}_{\alpha\sigma}\hat\Gamma{}^\sigma{}_{\mu\beta}=0\,.
\end{equation}

In these coordinates, since the parametrization of metric gives
\begin{equation}
  \cosh\hat\sigma = \hat g_{+-}/\sqrt{-\hat g} = 1
  \,\qquad
  \tanh\frac{\hat\sigma}2 = 0\,,
\end{equation}
the solution \eqref{eq:GeneralestW} can be written as simply as
\begin{equation}
  \hat W{}^\mu = \frac12\partial_\nu\hat g^{\mu\nu} + \frac{\varepsilon^{\mu\nu}}{2\sqrt{-\hat g}}\partial_\nu\hat r\,,
\end{equation}
Another simple identity holds
\begin{equation}\label{eq:ChiralDivergence}
  \varepsilon^{\alpha\beta}\partial_\beta\hat W{}_\alpha =
  -\frac{\sqrt{-\hat g}}{2}\hat g^{\alpha\beta}\partial_\alpha\partial_\beta \hat r\,.
\end{equation}
It should be stressed, that $\gamma$ in \eqref{eq:GammaParametrization} is an undefined function in isothermal coordinates. We assume that $\gamma$ and its derivatives are bounded.

Substituting the previous identities and setting $r=0$ for a moment, the improvement of EMT \eqref{eq:EMTExtraTerm} becomes
\begin{equation}
  \frac{\beta^2}{4}\hat T{}^{\mu\nu}_{\text{extra}}
  =
  (\hat g^{\mu\alpha}\partial_\alpha\rho)(\hat g^{\nu\beta}\partial_\beta\rho)
  -\frac12\hat g^{\mu\nu}(\hat g^{\alpha\beta}\partial_\alpha\rho\partial_\beta\rho)
  +\qty[\hat g^{\mu\nu}(\hat g^{\alpha\beta}\partial_\alpha\partial_\beta)-\hat g^{\mu\alpha}\hat g^{\nu\beta}\partial_\alpha\partial_\beta]\rho\,.
\end{equation}
Notice the close resemblance to the Liouville EMT\footnote{Equivalent situation appears for the Polyakov effective action, where $w(x)$ in the EMT follows the same structure, \citep{Jackiw_1995}.} \eqref{eq:CurvedEMTLiouville} upon taking $\beta=2$ and $m^2=0$. The ``field'' $\rho$ even transforms as a Liouville field under Weyl transformation \eqref{eq:RicciWeylGauging2D}
\begin{equation*}
  \rho\weyl \rho + \omega\,,
\end{equation*}
and the equation \eqref{eq:RicciScalarIsothermalCoordinates} replicates the Liouville equation of motion \eqref{eq:CurvedLiouvilleEquation}. However, it should not be forgotten that $\rho$ is part of the metric tensor $\hat g_{\mu\nu}$ and that $\rho$ is still a fixed background field. This is where the analogy with the dynamical Liouville scalar field stops.

The divergence \eqref{eq:EMTDiffeoBroken} of $T^{\mu\nu}_{\text{extra}}$ simplifies in the isothermal coordinates to
\begin{equation}\label{eq:EMTIsothermalDivergence}
    \beta^2\hat\nabla{}_\mu \hat T{}^{\mu\nu}_{\text{extra}}
    \frac{\varepsilon^{\nu\mu}}{\sqrt{-\hat g}}\hat g^{\alpha\beta}\partial_\alpha\partial_\beta \partial_\mu\hat r
    -2 \hat g^{\alpha\beta}\partial_\mu\qty(\hat r^{\mu\nu}\partial_\alpha\partial_\beta \hat r)
    + 2\partial_\mu \hat g^{\alpha\beta}\hat r^{\mu\nu}\partial_\alpha\partial_\beta \hat r\,.
\end{equation}
Not only it does not coincide with \eqref{eq:WilczekDiffeoAnomaly} or \eqref{eq:JackiwDiffeoAnomaly}, but it does vanish for $r$ satisfying
\begin{equation}\label{eq:RRestriciton}
  \hat\Box_M \hat r\equiv \eta^{\alpha\beta}\partial_\alpha\partial_\beta \hat r = 0\,,
\end{equation}
and $T^{\mu\nu}_{\text{extra}}$, thus, may produce a traceless EMT which also has a vanishing divergence in isothermal coordinates. Notice that
\begin{equation}
  \hat g^{\alpha\beta} \partial_\alpha\partial_\beta \hat r = \frac1{\sqrt{-g}}\hat\Box_M \hat r \,,
\end{equation}
and that for this choice, \eqref{eq:RRestriciton}, the right-hand side of \eqref{eq:ChiralDivergence} also vanishes
\begin{equation}\label{eq:ChiralDivergence2}
  \varepsilon^{\alpha\beta}\partial_\beta\hat W{}_\alpha = 0\,.
\end{equation}

Although, equation \eqref{eq:EMTIsothermalDivergence} looks like a covariant conservation for the choice \eqref{eq:RRestriciton}
\begin{equation}\label{eq:EMTIsothermalDivergenceRVanish}
  \hat\nabla{}_\mu \hat T{}^{\mu\nu}_{\text{extra}}\eval_{\hat\Box_M \hat r=0}
  = 0,
\end{equation}
it has to be emphasized that $T^{\mu\nu}_{\text{extra}}$ is not a tensor. Under a general coordinate transformation the \eqref{eq:EMTIsothermalDivergenceRVanish} does not have to necessarily hold. Therefore, to investigate diffeomorphism invariance one more step has to be taken and the transformation of $\hat\nabla_\mu \hat T{}^{\mu\nu}_{\text{extra}}$ has to be found. In the rest of the paper the choice \eqref{eq:RRestriciton} is assumed, since such $r$ may lead to simultaneous Weyl and diffeomorphism invariance.

\section{Breaking diffeomorphism invariance} \label{Sec:DiffeoBreaking}

The loss of diffeomorphism invariance is the loss of general covariance. Thus, the equations depend on the chosen coordinate frame and the results cannot be easily transformed between coordinates. Furthermore, the choice of the initial coordinate frame determines what physics the observer sees, hence it stands as a privileged frame. With the choice of isothermal coordinates, as initial coordinate frame, we might have just been lucky. The transformation of \eqref{eq:EMTDiffeoBroken} under a general coordinate transformation, or more precisely, of its non-tensorial part
\begin{equation}\label{eq:EMTDivergenceTransformationDeviation}
  \Delta \hat\nabla{}_\mu \hat T{}^{\mu\nu}_{\text{extra}}(x)
  \equiv
  \nabla'_\mu T'{}^{\mu\nu}_{\text{extra}}(x')
  -\pdv{x'^\nu}{x^\rho}\hat\nabla{}_\sigma \hat T{}^{\sigma\rho}_{\text{extra}}(x)\,,
\end{equation}
should be derived, to prove diffeomorphism (non-)invariance. This has to be, at least partially, expressible in terms of
$\Delta \hat W{}^{\mu}(x)
  \equiv
  W'^{\mu}(x')
  - (\partial x'^{\mu} / \partial x^\nu ) \hat W{}^\nu(x)
$.{color{blue}
For the infinitesimal diffeomorphism \eqref{eq:InfinitesimalDiffeo} the transformation of $W^\mu$ follows \eqref{eq:JackiwSolutionDiffeoTransf} and the non-tensorial part is
\begin{equation}
    \Delta \hat W{}^\mu
    =
    \frac{\varepsilon^{\mu\nu}}{2\sqrt{-\hat g}}\partial_\nu
    \qty(\partial_-f^- - \partial_+f^+)
    +
    \qty(
      \delta^\mu_\alpha - \partial_\alpha f^\mu
    )
    \frac{\varepsilon^{\alpha\beta}}{2\sqrt{-\hat g}}
    \partial_\beta \Delta r\, ,
\end{equation}
where $\Delta r$ was defined as the deviation of $r$ from the scalar transformation $\Delta r(x) \equiv r'(x') - r(x)$.

As $r(x)$ is only assumed to be Weyl scalar, \eqref{eq:WAmbiguity}, its transformation under diffeomorphism is not specified. To improve the transformation properties of $W^\mu$, $r(x)$ has to tackle the extra terms of \eqref{eq:JackiwSolutionDiffeoTransf}, which are proportional to $f^\mu$. It is, thus, natural to assume that $\Delta r$ contains at least the first power of $f^\mu$ in all of its terms, since finite terms (terms not containing $f^\mu$) in $\Delta r$ would necessarily inflate the deviation of $W^\mu$ from a contravariant vector. Under this assumption the previous equation is
\begin{equation}\label{eq:NonTensorialWTransf}
  \Delta \hat W{}^\mu
  =
  \frac{\varepsilon^{\mu\nu}}{2\sqrt{-\hat g}}\partial_\nu\,
  \partial_\alpha
  \qty(
    \frac{\varepsilon^{\alpha\beta}}{\sqrt{-\hat g}}f_\beta
    )
  +
  \frac{\varepsilon^{\mu\nu}}{2\sqrt{-\hat g}}
  \partial_\nu \Delta r
  \equiv
  \frac{\varepsilon^{\mu\nu}}{2\sqrt{-\hat g}}\partial_\nu \xi(f, r)\,,
\end{equation}
and $\Delta\hat W{}^\mu$ is proportional to the infinitesimal factor $f^\mu$.

Taking this as a starting point, it is clear that several terms in \eqref{eq:EMTDiffeoBroken} do not contribute to the infinitesimal transformation. In the isothermal coordinates and under the assumption \eqref{eq:RRestriciton}, $\hat\Box_M \hat r = 0$, the factors $(\cosh \hat \sigma-1)$ and $\varepsilon^{\alpha\beta}\partial_\beta W_\alpha$ separetely vanish. Terms in \eqref{eq:EMTDiffeoBroken} containing both factors effectively disappear since two vanishing factors are multiplied and only one at a time can be replaced by a factor containing $f^\mu$. The effective part of $\nabla_\mu T^{\mu\nu}_{\text{extra}}$ of concern in \eqref{eq:EMTDivergenceTransformationDeviation} is then
\begin{equation}
  \begin{split}
    \beta^2\nabla_\mu T^{\mu\nu}_{\text{extra}} \overset{\text{eff}}=& ~ 2\nabla_\rho\nabla^\rho W^\nu
    - RW^\nu
    + 4 \frac{\varepsilon^{\alpha\beta}}{\sqrt{-g}}
      \partial_\mu\qty(
        \partial_\beta W_\alpha r^{\mu\nu}
      )
    \\
    &  + 4 \frac{\varepsilon^{\alpha\beta}}{\sqrt{-g}}
    \partial_\beta W_\alpha
    g^{\mu\nu}\partial_\sigma g_{\lambda\mu}
    \qty(
      r^{\sigma\lambda}
      +\frac12\frac{\varepsilon^{\sigma\lambda}}{\sqrt{-g}}
    )
    \,.
  \end{split}
\end{equation}
}

The non-tensorial part of the transformation is
\begin{equation}
  \begin{split}
    \beta^2\Delta\hat\nabla{}_\mu\hat T{}^{\mu\nu}_{\text{extra}} =&~ 2\hat\nabla{}_\rho\hat\nabla{}^\rho \Delta\hat W{}^\nu
    - \hat R\Delta \hat W{}^\nu
    +4 \frac{\varepsilon^{\alpha\beta}}{\sqrt{-\hat g}}
      \partial_\mu\qty[
        \partial_\beta(\hat g_{\alpha\rho}\Delta \hat W{}^\rho)\hat r^{\mu\nu}
      ]
    \\
    &+4 \frac{\varepsilon^{\alpha\beta}}{\sqrt{-\hat g}}
    \partial_\beta\qty(\hat g_{\alpha\rho}\Delta \hat W{}^\rho)
    \hat g{}^{\mu\nu}
    \partial_\sigma \hat g_{\lambda\mu}
    \qty(
      \hat r^{\sigma\lambda}
      +\frac12
      \frac{\varepsilon^{\sigma\lambda}}{\sqrt{-\hat g}}
    )
    \,.
  \end{split}
\end{equation}

Using \eqref{eq:NonTensorialWTransf}, and carefully evaluating each term, it follows
\begin{equation}\label{eq:DivergenceEMTtransformation}
  \beta^2\Delta\hat\nabla{}_\mu \hat T{}^{\mu\nu}_{\text{extra}}
  =
  \frac{\varepsilon^{\nu\mu}}{\sqrt{-\hat g}}
  \hat g^{\alpha\beta}
  \partial_\mu\partial_\alpha\partial_\beta\xi
  -2\hat g^{\alpha\beta}\partial_\mu\qty(\hat r^{\mu\nu} \partial_\alpha\partial_\beta\xi)
  +2\hat r^{\mu\nu}\partial_\mu \hat g{}^{\alpha\beta}\partial_\alpha\partial_\beta\xi\,.
\end{equation}
Using Jackiw's choice \eqref{eq:JackiwSolution}, $r=0$, this simplifies significantly to
\begin{equation}\label{eq:rzerochoice}
    \beta^2\Delta\hat\nabla{}_\mu \hat T{}^{\mu\nu}_{\text{extra}}
    =
    \frac{\varepsilon^{\nu\mu}}{\sqrt{-\hat g}}
    \hat g^{\alpha\beta}
    \partial_\mu
      \partial_\alpha\partial_\beta
      \qty(\partial_-f^--\partial_+f^+)\,.
\end{equation}
This does not vanish for a general $f^\mu$, proving the loss of diffeomorphism invariance in the Weyl invariant formulation of Liouville theory \eqref{eq:RicciGaugedLiouville}, with a choice of Deser-Jackiw solution \eqref{eq:JackiwSolution}
\begin{equation} \label{eq:DeserJackiwLiouville}
  \mathcal A_{\scriptscriptstyle{LDJ}}[\Phi] = \int \dd[2]{x}\sqrt{-g}
  \qty(
    \frac{1}{2}g^{\mu\nu}\partial_\mu\Phi\partial_\nu\Phi
    - \frac{m^2}{\beta^2}e^{\beta\Phi}
    +\frac{1}{\beta}\Phi R
    +\frac{2}{\beta^2}g_{\mu\nu}W^\mu_{\scriptscriptstyle{DJ}} W^\nu_{\scriptscriptstyle{DJ}}) \,.
\end{equation}

Therefore for a general local solution of $2\nabla_\mu W^\mu = R$ the Weyl invariance is achieved at the cost of diffeomorphism invariance.

Whether a diffeomorphism can be restored through a suitable choice of $r$ is unclear to us. Hopefully the equations \eqref{eq:EMTIsothermalDivergence} and \eqref{eq:DivergenceEMTtransformation} may hint appropriate choices of $r$, as well as the fact, that $r$ needs to be Weyl scalar, $r\weyl r$, to preserve Weyl transformation of $W^\mu$ \eqref{eq:WeylGFWeylTransformation}.

One set of metric-based Weyl scalars is formed by $\sqrt{-g}g^{\alpha\beta}$ and $g_{\alpha\beta}/\sqrt{-g}$ as suggest by the parametrization \eqref{eq:MetricParametrization}. Moreover, in isothermal coordinates these are constants and any $r$ built out of them automatically satisfies \eqref{eq:RRestriciton}.

The loss of diffeomorphism, together with the non-trivial form of $T^{\mu\nu}_{\text{extra}}$, jeopardize the simple transformations between coordinate frames. Moreover, the complete lack of diffeomorphism hinders from using our results with isothermal coordinates as a midpoint, since infinitesimal transformation between the coordinate frames may not exist. The form of \eqref{eq:rzerochoice} suggests there might be a subset of diffeomorphisms which do not violate the general covariance. Such subset is of great importance since suitable choice of coordinate frame can reduce the degrees of freedom of the problem, as the illustrated choice of isothermal coordinates. A general two-dimensional metric $g_{\mu\nu}$ given by three independent functions $g_{tt}$, $g_{tx}=g_{xt}$ and $g_{xx}$ was reduced to a single function $\rho$ and the two degrees of freedom were removed by the diffeomorphism $x^\mu\goto \hat x^\mu(x)$.

Compared to the diffeomorphisms, the supplied Weyl invariance can reduce only one degree of freedom. Nonetheless, some theories allow keeping the Weyl invariance, while reducing the diffeomorphism invariance to invariance under area-preserving diffeomorphisms, \citep{Karakhanyan_1994, Jackiw_1995}. These are the diffeomorphisms \eqref{eq:InfinitesimalDiffeo} such that
\begin{equation}\label{eq:AreaPreservingDiffeo}
  \partial_\mu f^\mu = 0 \,.
\end{equation}
Such subset of diffeomorphisms can remove one degree of freedom, as $f^\mu$ may be written as $f^\mu = \varepsilon^{\mu\nu}\partial_\nu f$, where $f$ is an arbitrary function.

As a starting point to find the transformations which respect the tensorial behavior of $\hat\nabla_\mu \hat T{}^{\mu\nu}_{\text{extra}}$ we move to the flat limit by taking $\rho \goto 0$ in \eqref{eq:DivergenceEMTtransformation}
\begin{equation}\label{eq:FlatEMTDnontensorial}
    \beta^2\Delta\hat\nabla{}_\mu \hat T{}^{\mu\nu}_{\text{extra}}\eval_{\rho\goto0}
    =
    \varepsilon^{\nu\mu}
    \partial_\mu\Box_M\xi(r,f)
    -2\partial_\mu\qty( r^{\mu\nu} \Box_M\xi(r,f))\,.
\end{equation}
The function $\xi(r,f)$ depends only on $f^\mu$ and the exact dependence on $f^\mu$ is determined by the choice of $r$, but regardless of the choice, $f^\mu$ is accompanied by at least first order derivative in each term of $\xi(r,f)$. Thus, the form of \eqref{eq:FlatEMTDnontensorial} suggests that quadratic polynomial transformations
\begin{equation}
  f^\mu = a^{\mu}{}_{\alpha\beta}x^\alpha x^\beta + b^\mu{}_\alpha x^\alpha + c^\mu\,,
\end{equation}
which include the Poincaré transformations
\begin{equation}
  f^\mu = \omega^\mu{}_\nu x^\nu + c^\mu\,,
\end{equation}
do not violate the tensorial transformation of $\hat\nabla_\mu \hat T{}^{\mu\nu}_{\text{extra}}$.

From the linearity of $\xi(r,f)$ in $f^\mu$  and commutativity of partial derivatives it follows that for conformal transformations
\begin{equation*}
  \Box_M f^\mu = 0\,,
\end{equation*}
the right-hand side of the \eqref{eq:FlatEMTDnontensorial} also vanish.

\emph{We conclude that for infinitesimal conformal and Poincaré transformations in flat spacetime the extra term of the EMT,} $T^{\mu\nu}_{\text{extra}}$\emph{, is conserved, regardless of the choice of} $r$. Thus, the extra term does not violate the symmetries of the EMT in the flat limit.

It should be noted that the subset of conformal diffeomorphisms leaves the metric rescaled and their effect on metric is equivalent to Weyl transformations and compared to the area-preserving diffeomorphisms \eqref{eq:AreaPreservingDiffeo} it does not provide coordinates simplifying a general metric in already Weyl invariant theory. Therefore, we hope to find another transformations for which $\hat\nabla_\mu \hat T{}^{\mu\nu}_{\text{extra}}$ transforms tensorially. It is clear, that such additional transformations depend on the choice of $r$.

For the choice $r=0$ \eqref{eq:FlatEMTDnontensorial} simplifies to
\begin{equation}
  \beta^2\Delta\hat\nabla{}_\mu \hat T{}^{\mu\nu}_{\text{extra}}\eval_{\rho\goto0}
  =
  \varepsilon^{\nu\mu}
  \partial_\mu\Box_M
  \qty(
    \varepsilon^{\alpha\beta}
    \partial_\alpha f_\beta
  )\,.
\end{equation}
This immediately shows that the diffeomorphisms \eqref{eq:AreaPreservingDiffeo} do not respect tensorial transformations. Whether there exists a choice of $r$ such that the area-preserving diffeomorphisms are satisfied is unclear to us. However, a transformation
\begin{equation}
  f_\mu = \partial_\mu f\,,
\end{equation}
where $f$ is an arbitrary function, does respect the tensorial transformations.
It should be noted that this is rather a hint and more work has to be done to generalize this transformation back to curved background in such a way that $ W{}_{\scriptscriptstyle{DJ}}^\mu(x)$ transforms vectorially.

~

As demonstrated, the Weyl invariance produces traceless EMT
\begin{equation}\label{eq:TotalWeylEMT}
  \tilde T{}^{\mu\nu}\equiv T^{\mu\nu}_{\scriptscriptstyle{L}} + T^{\mu\nu}_{\text{extra}}\,.
\end{equation}
That means the center manifested in the ``anomalous'' trace vanishes. Nonetheless, the center is still present in the transformation of the curved space EMT \eqref{eq:TotalWeylEMT}, as it was in the transformation of the flat spacetime version of the EMT, \eqref{eq:InfinitesimalEMTtransformation}. For the infinitesimal diffeomorphism \eqref{eq:InfinitesimalDiffeo}, $x^\mu\goto x^\mu - f^\mu(x)$,
the non-tensorial transformation of the extra term $T^{\mu\nu}_{\text{extra}}$ is
\begin{equation}\label{eq:CurvedAnomalousTransformationEMT}
  \begin{split}
    \beta^2\Delta\hat T{}^{\mu\nu}_{\text{extra}}(x)\equiv&~  \beta^2 \left( \hat  T'^{\mu\nu}_{\text{extra}}(x')-\pdv{x'^\mu}{x^\rho}\pdv{x'^\nu}{x^\sigma}T^{\rho\sigma}_{\text{extra}}(x) \right)
    \\
    =& ~
    \qty(
      \hat g^{\mu\alpha}\frac{\varepsilon^{\nu\beta}}{\sqrt{-\hat g}}
      + \hat g^{\nu\alpha}\frac{\varepsilon^{\mu\beta}}{\sqrt{-\hat g}}
      )
    \partial_\alpha \partial_\beta \xi
    \\
    &+
    (\hat g^{\mu\nu}\hat g^{\alpha\beta} - \hat g^{\mu\beta}\hat g^{\nu\alpha} - \hat g^{\mu\alpha}\hat g^{\nu\beta})
    \partial_\alpha \hat r\partial_\beta \xi
    -2\hat r^{\mu\nu}\hat g^{\alpha\beta}\partial_\alpha\partial_\beta \xi \,,
  \end{split}
\end{equation}
which reduces for the choice $r=0$ to
\begin{equation}
    \beta^2\Delta\hat T{}^{\mu\nu}_{\text{extra}}(x)
    =
    \qty(
      \hat g^{\mu\alpha}\frac{\varepsilon^{\nu\beta}}{\sqrt{-\hat g}}
      + \hat g^{\nu\alpha}\frac{\varepsilon^{\mu\beta}}{\sqrt{-\hat g}}
      )
    \partial_\alpha \partial_\beta \xi
\end{equation}
Assuming conformal diffeomorphisms and taking the flat limit\footnote{Let us remind here that $\varepsilon^{-+}=+1$.} we have
\begin{equation}
  \Delta\hat T{}^{\pm\pm}_{\text{extra}}(x)\eval_{\rho\goto0}
  = -\frac2{\beta^2} \partial^3_\mp f^\mp
  \,.
\end{equation}
Thus, the full improved EMT \eqref{eq:TotalWeylEMT} transforms non-tensorially
since
\begin{equation}
  \Delta \tilde T{}^{\mu\nu} = \Delta T^{\mu\nu}_{\scriptscriptstyle{L}} + \Delta T^{\mu\nu}_{\text{extra}} = \Delta T^{\mu\nu}_{\text{extra}}\,,
\end{equation}
and it follows that
\begin{equation}\label{eq:CurvedCenter}
  \Delta \hat{\tilde{T}} {}^{\pm\pm}(x)\eval_{\rho\goto0}
  = -\frac2{\beta^2} \partial^3_\mp f^\mp
  \,.
\end{equation}

The measure of non-tensoriality $\Delta$ is a universal operation and we can apply it on the EMT of flat Liouville theory $\tilde \Theta_{\mu\nu}$, \eqref{eq:FlatLiouvilleEMTimproved},
\begin{equation}\label{eq:FlatCenter}
  \Delta \tilde\Theta_{\pm\pm}
  \equiv
  \tilde\Theta'_{\pm\pm}(x')  - \pdv{x^\mu}{x'^\pm}\pdv{x^\nu}{x'^\pm}\tilde\Theta_{\mu\nu}(x)
  = -\frac2{\beta^2} \partial^3_\pm f^\pm
  \,,
\end{equation}
where the last equality follows from \eqref{eq:InfinitesimalEMTtransformation}. To make contact between the two EMTs, just notice $\Theta^{\pm\pm} = \Theta_{\mp\mp}$. Thus \eqref{eq:CurvedCenter} is exactly the extra anomalous part of conformal transformation of the flat Liouville EMT $\tilde \Theta_{\mu\nu}$, \eqref{eq:FlatCenter},
\begin{equation}
  \Delta\hat{\tilde{T}}{}^{\pm\pm}\eval_{\rho\goto0}
  =
  \Delta \tilde\Theta{}^{\pm\pm}
  \,,
\end{equation}
with center given by $c = 48 \pi/\beta^2$.

This way the center is removed from the trace of the EMT, restoring Weyl invariance, but it appears in its transformations, breaking general covariance.


\section{Conclusions}

In the framework of the general results on conformal symmetry of \cite{WeylGauging}, based on the assumption that diffeomorphism and Weyl invariances for classical field theories can be made to hold together, and motivated by Jackiw's conjecture \citep{WeylvsLiouville}, proved in \cite{Letter1HamanIorio2023}, here we enlarged the scope of \cite{Letter1HamanIorio2023} and gave a detailed explanation of many of the results reported in that paper.

We did so by searching for the classical central charges and by studying the implications of their presence in a classically anomalous theory, that is a theory with genuine central charge in the algebra of Noether charges. We then studied conformal transformations in two dimensions, where they satisfy the Witt algebra. The emergence of a center $c=48\pi/\beta^2$ of the Virasoro algebra of conformal Noether charges for Liouville  theory in Minkowski spacetime is demonstrated in detail. Such center is responsible for an anomalous transformation of the EMT under the conformal transformations. Moreover, the form of the anomalous transformation is the same as an anomalous transformation of an EMT known in conformal \emph{quantum} field theories.

We then moved to Liouville theory in curved spacetime. The diffeomorphism invariant version of the theory is Weyl-gauged, leading to diffeomorphism and Weyl invariant Liouville theory with a Weyl-gauge field, $W^\mu$. The condition \eqref{eq:WeylRicci2DIdentification}, $2\nabla_\mu W^\mu=R$, allows to trade the Weyl-gauge field for metric-dependent terms, introducing a non-minimal coupling to the scalar curvature, $R$, and a purely geometrical improvement, independent from the dynamical Liouville field. This trade-off (Ricci gauging, in the language of \cite{WeylGauging}) formally seems to keep the diffeomorphism and Weyl invariance. In fact, it is not so, as the solution to condition \eqref{eq:WeylRicci2DIdentification} is \textit{either non-local} (leading to the Polyakov effective action) or it leads to a quantity that \textit{does not transform as a contravariant vector}. Therefore, if one wants  to preserve Weyl invariance, either locality or diffeomorphism invariance is lost. Notice that the very same geometrical improvement was found in \citep{WeylvsLiouville}, where Weyl invariant Liouville theory was constructed from a dimensional limit.

Choosing the local, non-tensorial, Deser-Jackiw solution to the condition, \eqref{eq:WeylRicci2DIdentification}, we proceeded with calculating the extra contributions to the EMT stemming from the geometrical improvement. Since diffeomorphism and Weyl invariances of a theory are translated to properties of its EMT, the tracelessness, implied by the latter invariance, is demonstrated for an arbitrary metric in Weyl invariant Liouville theory. Following with explicit calculations, in isothermal coordinates, we showed that for a general local solution to the condition \eqref{eq:WeylRicci2DIdentification} the EMT is not conserved because the transformation
\begin{equation*}
  \beta^2\Delta\hat\nabla{}_\mu \hat T{}^{\mu\nu}_{\text{extra}}
  =
  \frac{\varepsilon^{\nu\mu}}{\sqrt{-\hat g}}
  \hat g^{\alpha\beta}
  \partial_\alpha\partial_\beta
  \partial_\mu
  \qty(\partial_-f^--\partial_+f^+)
  \,,
\end{equation*}
does not vanish for a general $f^\mu$. This implies the loss of diffeomorphism invariance for Weyl invariant Liouville theory. This proves Jackiw's conjecture \citep{WeylvsLiouville}.

Deser-Jackiw solution to the condition \eqref{eq:WeylRicci2DIdentification} contains an undetermined function $r$. This ambiguity and its relation to the preservation of diffeomorphism invariance were studied. In \eqref{eq:EMTIsothermalDivergence} and \eqref{eq:DivergenceEMTtransformation} we explicitly derived the necessary conditions $r$ must satisfy for diffeomorphism invariance to hold. However, we were not able to find a solution to the conditions and whether diffeomorphism invariance is achievable, through a particular choice of $r$, is unclear. The formula \eqref{eq:DivergenceEMTtransformation} also shows how the choice of $r$  is related to the preserved subset of diffeomorphisms. Regardless of the choice, though, Poincaré and global conformal symmetries are preserved, in the flat limit. This substantiates the results of \cite{WeylGauging} on scale invariance and full conformal invariance of Liouville theory in flat space.

In parallel, we tracked the presence of the central charge in curved background. Starting with the commonly used Liouville theory, which is diffeomorphism invariant and non-minimally coupled to curvature, the trace of its EMT provides a classical instance of the trace anomaly formula, since it is proportional only to the center $c$ and curvature $R$. When the theory is made Weyl invariant the trace vanishes. The center does not disappear from the theory, though, but rather moves into the transformation of the EMT \eqref{eq:CurvedAnomalousTransformationEMT}, which, in the flat limit, gives back the anomalous conformal transformation of the EMT
\begin{equation*}
  \Delta\hat T{}^{\pm\pm}_{\text{extra}}(x)\eval_{\rho\goto0}
  =-\frac{c}{24 \pi} \partial^3_\mp f^\mp.
\end{equation*}

Although the value of the central charge of Liouville theory changes in the process of quantization due to the additional contributions from normal ordering, the forms of Virasoro algebra, anomalous transformation of the EMT and the trace anomaly remain the same. Liouville theory, thus, may provide a starting point for further studies to discern the effects of central charge from the effect of quantization.

In our study we focused mostly on Liouville theory. Whether it is possible, in general, to derive trace or diffeomorphism anomalies directly from Virasoro algebra in a classical system is still an open question. The equivalence of central charge and trace anomaly, in quantum theories is established by using methods of path integration, which are unavailable for classical theories. Nonetheless, it is reasonable to suppose that the similarity of the results, in the quantum case and in the classical case, is not coincidental. Further work on the central charge of Virasoro algebra in classical theories could give a definite answer.

A last direction for further research that we would like to mention here, is the investigation of the connection of the anomalous transformation of the EMT with classical analog of Unruh and Hawking effects.

\section*{Acknowledgments}

We gladly acknowledge support from Charles University Research Center (UNCE/SCI/013).

\bibliographystyle{apsrev4-2}

%


\newpage
\appendix
\section{Passive and active transformations}\label{app:NoetherTheorem}

For spatiotemporal symmetries we can take the \emph{passive transformations} point of view: If $x^\mu$ is a coordinate system and $f^\mu(x)$ a smooth vector field over the spacetime, starting at point $p$, with coordinates $x^\mu(p)\equiv p^\mu$, one can move along the flow lines of the vector field $f^\mu$ by an infinitesimal distance parametrized by $\epsilon$ to a new point $q$ with coordinates $x^\mu(q)\equiv q^\mu$
\begin{equation}\label{eq:CoordinateTransformation}
  p^\mu\goto q^\mu=p^\mu + \epsilon f^\mu(p) \,.
\end{equation}
In differential geometry this is known as the \emph{flow generated by the vector field $f^\mu$} and it is denoted by $\phi_\epsilon$, \citep{GeometryTopologyPhysics_Nakahara}. This can be then recast into the coordinate-free identification $\phi_\epsilon(p)=q$.

Instead of moving the observer in a given direction, the same result can be achieved by transporting the fields in the inverse direction, i.e. from the point $q$ to the point $p$. That is the \emph{active transformations} point of view. The point $q$ in a new coordinate frame, $x'^\mu$, can be chosen to have the same coordinate values as the point $p$ in the original frame, $x^\mu$, that is, $x'^\mu(q)=x^\mu(p)$ or, equivalently,
\begin{equation}\label{eq:PullbackCondition}
  q'^\mu=p^\mu \,.
\end{equation}
Solving this equation up to the first order gives $x'^\mu(p) = x^\mu(p)-\epsilon f^\mu(p)$.

E.g., taking a vector field $V$, in the coordinate frame $x^\mu$, and applying coordinate transformation rules at point $q$ for vectors one obtains
\begin{equation}\label{eq:PullbackVectorComponents}
  V'^\alpha(x'(q))=\pdv{x'^\alpha(q)}{x^\beta(q)} V^\beta(q).
\end{equation}
Using the condition \eqref{eq:PullbackCondition}, $V'^\alpha(x'(q))=V'^\alpha(x(p))$, which introduces a ``new'' vector field $V'$, with components $V'^\alpha$ in the \emph{original} coordinate frame $x^\mu$ and equation \eqref{eq:PullbackVectorComponents} as its definition.

The new field $V'$ can be compared with the original field $V$ at the point $p$ by expanding the equation \eqref{eq:PullbackVectorComponents}
\begin{equation}\label{eq:TransformationVectorField}
    V'^\alpha(p) = \pdv{x'^\alpha(q)}{x^\beta(q)} V^\beta(q)
    = V^\alpha(p)+\epsilon f^\beta\partial_\beta V^\alpha(p) - \epsilon \partial_\beta f^\alpha(p) V^\beta(p) \,,
\end{equation}
where in the last equation everything was expanded up to the first order in $\epsilon$.

The solution to the coordinate transformation \eqref{eq:PullbackCondition} is the \emph{Lie transport} along the flow lines of $f^\mu$ from point $q$ back to the point $p$. In general relativity this backward transport of field is called \emph{pullback}, as it is ``pulling'' the field back opposed to the direction of flow $\phi_\epsilon$, and it is denoted as $\phi^*_\epsilon$, \citep{CarrollRelativity}. Within this more abstract notation the equation \eqref{eq:PullbackVectorComponents} can be rewritten as
\begin{equation}\label{eq:PullbackVectorField}
  V'(p)=\phi^*_\epsilon[V(\phi_\epsilon(p))],
\end{equation}
which consists of two purely geometrical operations, a flow from the point $p$ to the point $\phi_\epsilon(p)$ and a pull of the field $V$ back to the point $p$. This is a \emph{diffeomorphism}, and it is an active transformation, since the fields are transformed, rather than the observer.

For transformations not leaving or changing the spacetime, the equation \eqref{eq:PullbackVectorField} holds for a general tensor field $T$ as
\begin{equation}\label{eq:PullbackTensorField}
  T'(p)=\phi^*_\epsilon[T(\phi_\epsilon(p))] \,.
\end{equation}
Again, starting with a field $T$ a new field $T'$ was obtained such that its domain is the same as of the original tensor field.

\section{On singular Lagrangians}\label{app:SingularLagrangian}

The case of Lagrangian density linear in the first order time derivatives of fields (linear in generalized velocities) is often called singular. Singular in a way that\footnote{There is no summation over the index $i$ in the formula.},
\begin{equation}
  \pdv{\mathcal{L}}{\dot\Phi{\vphantom{\Phi}}^i}{\dot\Phi{\vphantom{\Phi}}^i} = 0\,.
\end{equation}
In such situation the standard Legendre transformation is problematic, since it is not generally possible to invert the definition of conjugated momenta to obtain $\dot\Phi{}^i$ and switch to Hamiltonian formalism. This leads to constrained systems and the way to deal with them was found by Paul Dirac,\citep{Dirac1950GeneralizedHamiltonian}, by a generalization of Poisson brackets giving Dirac brackets.

Examples of ``singular'' Lagrangians may be the Schr{\"o}dinger Lagrangian or Dirac Lagrangian, where in both theories a Hamiltonian and Poisson brackets (or equivalently commutators) are well-defined, and the theories are not considered being singular.

For a subset of singular Lagrangians, there exists a complementary ``light-weight'' method to Dirac's approach introduced by Faddeev and Jackiw, \citep{Jackiw1993ConstrainedQuantization}. This method, if possible,  derives Poisson brackets. Brief introduction to the method in the context of the field theory is given here, to find Poisson brackets for light-cone (and similar) Lagrangians. It is assumed Lagrangian depends on one scalar field $\phi$ for simplicity, however, the method can be generalized to other fields and for multiple fields.

A general form of a Lagrangian linear in $\dot\phi$ is
\begin{equation}
  L[\phi,\dot\phi](t) = \int\dd[d-1]{\vb{x}}\qty(F[\phi](\vb{x},t)\dot\phi(\vb{x},t)-V[\phi](t))\,,
\end{equation}
where $F[\phi](\vb{x},t)$ is some non-local functional. A naive Legendre transformation to obtain Hamiltonian gives
\begin{equation}
  H = \fdv{L}{\dot\phi}\dot\phi - L = V\,.
\end{equation}
Therefore, the Lagrangian can be rewritten as
\begin{equation}
  L[\phi,\dot\phi](t) = \int\dd[d-1]{\vb{x}}\qty(F[\phi](\vb{x},t)\dot\phi(\vb{x},t)-H[\phi](t))\,.
\end{equation}

Since Poisson brackets encompass the equations of motion, they have to be properly derived to trace carefully the connection between Poisson brackets and equations of motion for the case of a singular Lagrangian. The non-local functional $F[\phi](\vb x,t)$ forbids the use of field theoretic generalization of Euler-Lagrange equation and the equations of motion have to be derived variationally
\begin{equation}
  \dv{t}\fdv{L[\phi,\dot\phi](t)}{\dot\phi(\vb{x},t)}-\fdv{L[\phi,\dot\phi](t)}{\phi(\vb{x},t)}=0\,.
\end{equation}
A careful use of the chain rule for functional derivatives gives
\begin{equation}
  \dv{t}F[\phi](\vb{x},t)-\int\dd[d-1]{\vb{y}}\dot\phi(\vb{y},t)\fdv{F[\phi](\vb{y},t)}{\phi(\vb{x},t)}+\fdv{H[\phi](t)}{\phi(\vb{x},t)}=0\,.
\end{equation}
Expanding the total time derivative leads to
\begin{equation}
  \fdv{H[\phi](t)}{\phi(\vb{x},t)} = \int\dd[d-1]{\vb{y}}\dot\phi(\vb{y},t)\qty(\fdv{F[\phi](\vb{y},t)}{\phi(\vb{x},t)}-\fdv{F[\phi](\vb{x},t)}{\phi(\vb{y},t)})-\pdv{F[\phi](\vb{x},t)}{t}\,.
\end{equation}

In case of Lagrangians independent of explicit time dependence the last term, $\pdv*{F}{t}$, is zero, and since the scope of the paper is in specific Lagrangians without explicit time dependence, this term is omitted in the following. Thus,
\begin{equation}
  \fdv{H[\phi](t)}{\phi(\vb{x},t)} = \int\dd[d-1]{\vb{y}}\dot\phi(\vb{y},t)\Xi(\vb{x},\vb{y})\,,
\end{equation}
where
\begin{equation}
  \Xi(\vb{x},\vb{y})\equiv \fdv{F[\phi](\vb{y},t)}{\phi(\vb{x},t)}-\fdv{F[\phi](\vb{x},t)}{\phi(\vb{y},t)}\,.
\end{equation}

The crucial point is to realize, that by finding a function $\Xi^{-1}(\vb{x},\vb{y})$, inverse to $\Xi(\vb{x},\vb{y})$,
\begin{equation*}
  \int \dd[d-1]{\vb{y}}\Xi(\vb{x},\vb{y})\Xi^{-1}(\vb{y},\vb{z}) = \int \dd[d-1]{\vb{y}}\Xi^{-1}(\vb{x},\vb{y})\Xi(\vb{y},\vb{z}) = \delta(\vb{x}-\vb{z})\,,
\end{equation*}
the previous equation can be turned into
\begin{equation}\label{eq:TimeEvolutionSingular}
  \dot\phi(\vb{x},t) = \int \dd[d-1]{\vb{y}} \Xi^{-1}(\vb{x},\vb{y})\fdv{H[\phi](t)}{\phi(\vb{y},t)}\,.
\end{equation}

In Hamiltonian formalism the time evolution of a field is given by Poisson brackets of that field and Hamiltonian
\begin{equation}\label{eq:TimeEvolutionHamiltonian}
  \dot\phi = \pb{\phi}{H}\,.
\end{equation}
Thus, the right-hand side of \eqref{eq:TimeEvolutionSingular} contains the information about Poisson brackets. Expanding $H[\phi]$ in a functional Taylor series, it can be shown
\begin{equation}\label{eq:PBofFunctional}
  \pb{\phi(\vb{x},t)}{H[\phi](t)} = \int\dd[d-1]{\vb{y}}\pb{\phi(\vb{x},t)}{\phi(\vb{y},t)}\fdv{H[\phi](t)}{\phi(\vb{y},t)}\,.
\end{equation}
Comparing the equations \eqref{eq:TimeEvolutionSingular}, \eqref{eq:TimeEvolutionHamiltonian} and \eqref{eq:PBofFunctional} it immediately follows
\begin{equation}
  \pb{\phi(\vb{x},t)}{\phi(\vb{y},t)} = \Xi^{-1}(\vb{x},\vb{y})\,.
\end{equation}
Poisson brackets are, thus, defined by the $\Xi^{-1}$ and the relation
\begin{equation}
  \pb{B[\phi](t)}{C[\phi](t)} = \int \dd[d-1]{\vb{x}}\dd[d-1]{\vb{y}}\fdv{B[\phi](t)}{\phi(\vb{x},t)}\Xi^{-1}(\vb{x},\vb{y})\fdv{C[\phi](t)}{\phi(\vb{y},t)}\,,
\end{equation}
where $B[\phi](t)$ and $C[\phi](t)$ are functionals of $\phi$.

\subsubsection*{Light-cone Lagrangian}
A common scenario with a Lagrangian linear in $\dot\phi$ is the light-cone coordinate version of a theory. Such scenario is frequent throughout the given text, therefore, few words should be given. Transformation into light-cone coordinates does not introduce anything new, so calling such a Lagrangian singular may seem odd and if the standard coordinates allow Hamiltonian formulation, it is not expected that a simple coordinate transformation might lead to the loss of the Hamiltonian formulation.

The analysis restricts here to a two-dimensional case. Nonetheless, treatment of higher dimensional cases is of no difference. In the case of light-cone scalar field theory, the kinetic term turns out to be
\begin{equation}
  \mathcal{L}_{(kin)} = \partial_+\phi(x^+,x^-)\partial_-\phi(x^+,x^-)\,.
\end{equation}
The ``time'' variable can be chosen at will, since there is no mathematical difference between $x^+$ and $x^-$ coordinates as the metric does not assign a special signature to one. This can be seen in Appendix \ref{app:EuclideMinkowski}. The time variable is chosen to be $x^+$, then the solution is identical to solving the problem for
\begin{equation}
  \begin{split}
    L[\phi,\dot\phi](t) =& \int\dd{x}\dot\phi(x,t)\partial_x\phi(x,t)\,,
    \\
    F[\phi](x,t) =& \int\dd{y}\partial_y\phi(y,t)\delta(y-x) = -\int\dd{y}\phi(y,t)\partial_y\delta(y-x)\,,
    \\
    \Xi(x,y) =& -\partial_x\delta(y-x)+\partial_y\delta(y-x) = 2\delta'(y-x)\,.
  \end{split}
\end{equation}

Inverting  $\Xi(x,y)$ gives
\begin{equation}
  \Xi^{-1}(x,y) = -\frac14\sgn(x-y)\,,
\end{equation}
where $\sgn(x)$ is the sign function
\begin{equation}
  \sgn(x) =
  \begin{cases}
    1 & x > 0\,,
    \\
    0 & x = 0\,,
    \\
    -1 & x < 0\,.
  \end{cases}
\end{equation}

Thus, Poisson brackets in the light-cone coordinates are
\begin{equation}
  \pb{\phi(x^+,x^-)}{\phi(x^+,y^-)}=-\frac14\sgn(x^--y^-)\,.
\end{equation}
In the same manner Poisson brackets for fixed $x^-$ are derived to be
\begin{equation}
  \pb{\phi(x^+,x^-)}{\phi(y^+,x^-)}=-\frac14\sgn(x^+-y^+)\,.
\end{equation}

Both Poisson brackets, for equal $x^+$ and for equal $x^-$, have to be defined because for some theories there might be a split of fields into subset of fields constant under the change of $x^+$ and subset of fields constant under the change of $x^-$. In such case both Poisson brackets are needed to get the full description.

\section{Euclide vs Minkowski in two dimensions}\label{app:EuclideMinkowski}

The conformal Killing equation \eqref{eq:2DConfKilling}
\[
  \partial_\mu f_\nu + \partial_\nu f_\mu = g_{\mu\nu}\partial_\rho f^\rho \,,
\]
holds for both signatures of the metric, the Euclidean, customarily used in conformal field theory\footnote{It is so, for its intrinsic connection with complex plane, allowing to use the results of complex analysis.} and the Minkowski. They are equivalent, but the details of this equivalence are hidden in the complexification of the coordinates and subsequent Wick rotation, which is discussed later here. In this paper we consider both signatures, thus the following table is most useful.

\begin{table}[H]
  \caption{Notations for Minkowski and Euclidean spacetimes}
  \label{tab:MinkowskiEuclideanSpaces}
  \centering
  \begin{tabular}{l @{\hskip 0.4in} l @{\hskip 0.4in} l}
    \toprule
                  & \textbf{Minkowski}       & \textbf{Euclidean}
    \\
    \midrule
    $\mu$         & $t,x$           & $0, 1$
    \vspace{0.1in}
    \\
    $x^\mu$       & $(t,x)$         & $(x^0, x^1)$
    \vspace{0.1in}
    \\
    $g_{\mu\nu}$  & $\eta_{\mu\nu}=
                      \begin{pmatrix}
                        1 & 0 \\
                        0 & -1
                      \end{pmatrix}$
                                    &   $\delta_{\mu\nu}=
                                          \begin{pmatrix}
                                            1 & 0\\
                                            0 & 1
                                          \end{pmatrix}$
    \vspace{0.1in}
    \\
    $\varepsilon^{\mu\nu}$
                  & $\varepsilon^{tx} = +1$
                                    &   $\varepsilon^{01} = +1$
    \vspace{0.1in}
    \\
    \multirow{2}{*}{Killing eq.}
                  &  $\partial_x f^x = \partial_t f^t$
                                    & $\partial_0 f^1 = -\partial_1 f^0$
    \\
                  & $\partial_x f^t = \partial_t f^x$
                                    & $\partial_0f^0=\partial_1f^1$
    \vspace{0.1in}
    \\
    $\partial_\nu\partial^\nu f_\mu =0$
                  & $\Box_{\text M} f_\mu = (\partial^2_t-\partial^2_x)f_\mu=0$
                                    & $\Box_{\text E} f_\mu=(\partial_0^2+\partial_1^2)
                                      f_\mu=0$
    \\
    \bottomrule
  \end{tabular}
\end{table}

Of course, the main difference between the two signatures is in the way the time coordinate is dealt with. While for Minkowski, one can naturally discern time, for the Euclidean both coordinates are equal on all footings. As a consequence, the conformal Killing equation turns into the \textit{wave equation} in Minkowski spacetime, while it is the \textit{Cauchy-Riemann conditions} for a function to be holomorphic (or, equivalently, the \textit{Laplace equation}) in the Euclidean space.

Both the wave equation and the Laplace equation, are better solved by introducing a new coordinates: the light-cone coordinates, for the wave equation, and the complex plane, for the Laplace equation. After the transformations both metrics attain the same form, and so does the conformal Killing equation, as illustrated in the table \ref{tab:LightConeComplexSpaces} below.

\begin{table}[h]
  \caption{Notations for light-cone and complex plane variables}
  \label{tab:LightConeComplexSpaces}
  \centering
  \begin{tabular}{l @{\hskip 0.4in} l @{\hskip 0.4in} l}
    \toprule
                  & \textbf{Light-cone}       & \textbf{Complex}
    \\
    \midrule
    $\mu$         & $-,+$           & $z, \bar z$
    \vspace{0.1in}
    \\
    \multirow{3}{*}{$x^\mu$}       & $(x^-,x^+)$     & $(z, \bar z)$
    \\
    & $x^-=\frac{1}{\sqrt2}(t- x)$
    & $z=\frac{1}{\sqrt{2}}(x^0+ix^1)$
    \\
    & $x^+=\frac{1}{\sqrt2}(t+ x)$
    & $\bar z=\frac{1}{\sqrt{2}}(x^0-ix^1)$
    \vspace{0.1in}
    \\
    \multirow{2}{*}{$\partial_\mu$}
    & $\partial_-=\frac1{\sqrt2}(\partial_t-\partial_x)$
    & $\partial\equiv\partial_z = \frac1{\sqrt2}(\partial_0-i\partial_1)$
    \\
    & $\partial_+=\frac1{\sqrt2}(\partial_t+\partial_x)$
    & $\bar\partial\equiv\partial_{\bar z} = \frac1{\sqrt2}(\partial_0+i\partial_1)$
    \vspace{0.1in}
    \\
    $g_{\mu\nu}$  & $\eta_{\mu\nu}=
                      \begin{pmatrix}
                        0 & 1 \\
                        1 & 0
                      \end{pmatrix}$
                                    &   $\delta_{\mu\nu}=
                                          \begin{pmatrix}
                                            0 & 1\\
                                            1 & 0
                                          \end{pmatrix}$
    \vspace{0.1in}
    \\
    $\varepsilon^{\mu\nu}$
                  & $\varepsilon^{-+} = +1$
                                    &   $\varepsilon^{z\bar z} = +1$
    \vspace{0.1in}
    \\
    \multirow{2}{*}{Killing eq.}
    &  $\partial_+ f^- = 0$
    & $\partial f^{\bar z} \equiv \partial \bar f = 0$
    \\
    & $\partial_- f^+ = 0$
    & $\bar \partial f^z\equiv\bar\partial f=0$
    \vspace{0.1in}
    \\
    $\partial_\nu\partial^\nu f_\mu =0$
    & $2\partial_+\partial_- f_\mu =0$
    & $2\partial\bar\partial f_\mu=0$
    \\
    \bottomrule
  \end{tabular}
\end{table}

The solutions to the conformal Killing equation split into two independent parts. The right-going, $f^-= f^-(x^-)$, and the left-going, $f^+= f^+(x^+)$, waves in the light-cone reformulation and the holomorphic, $f= f(z)$, and the antiholomorphic, $\bar f = \bar f(\bar z)$, complex functions in the complex space. This split hints that a theory allowing the conformal symmetry is separable into right-going and left-going, respectively holomorphic and antiholomorphic, subspaces.

Notice the chosen order of indices ($-, +$) for the light-cone coordinates in \ref{tab:LightConeComplexSpaces}. The order of indices determines the Levi-Civita symbol $\varepsilon^{\mu\nu}$. This choice, in particular, ensures that the transformation matrix
\begin{equation}
  S^\mu{}_\nu \equiv \pdv{x'^\mu}{x^\nu} = \frac{1}{\sqrt{2}}
  \begin{pmatrix}
    1 & -1 \\
    1 & 1
  \end{pmatrix}\,,
\end{equation}
has a positive and unit determinat, $\det S^\mu{}_\nu = +1$. Thus, tensor densities, e.g. $\sqrt{-g}$ and $\varepsilon^{\mu\nu}$, do not get any extra factor under this coordinate change.

\begin{figure}[ht]
  \centering
  \begin{subfigure}{.5\textwidth}
    \centering
    \begin{tikzpicture}
      \def\cylrad{1} 
      \def\cylcirc{2*3.14*\cylrad}
      \def\cylht{5} 
      \draw[ultra thick]
        \foreach \x in {0, \cylcirc} {(\x,0) -- (\x, \cylht)}
      ;
      \node[left] at (0, \cylht/2) {\rotatebox{90}{periodic boundary}}
      ;
      \draw[red, thick]
        \foreach \y in {0, \cylht/2} {(0, \y) -- (\cylcirc, \y+\cylht/2)}
      ;
      \node[right] at (\cylcirc*0.6, \cylht*0.78) {\color{red} $x^-=\text{const.}$}
      ;
      \draw[blue, thick]
        \foreach \y in {0, \cylht/2} {(\cylcirc, \y) -- (0, \y+\cylht/2)}
      ;
      \node[right] at (\cylcirc*0.2, \cylht*0.95) {\color{blue} $x^+=\text{const.}$}
      ;
      \fill[gray, opacity=0.2]
        (\cylcirc, \cylht/2) -- (\cylcirc/2, \cylht*3/4) -- (0, \cylht/2)
        -- (\cylcirc/2, \cylht/4) -- (\cylcirc, \cylht/2)
      ;
      \draw[<->]
        (0, 1.1*\cylht) -- (\cylcirc, 1.1*\cylht) node [midway, above] {$L$}
      ;
      \draw[fill]
        (\cylcirc/2, \cylht/4) circle [radius=1pt] node[below] {$x_0$}
      ;
      \draw[thick, ->] (\cylcirc*0.07,-0.05*\cylht) 
      -- (\cylcirc*0.07,-0.05*\cylht+1) node[anchor=north east] {$t$}
      ;
      \draw[thick, ->] (\cylcirc*0.07,-0.05*\cylht) %
      -- (\cylcirc*0.07+1.5,-0.05*\cylht) node[anchor=north] {$x$}
      ;
    \end{tikzpicture}
  \end{subfigure}%
  \begin{subfigure}{.5\textwidth}
    \raggedleft
    \begin{tikzpicture}[
      x=15mm,
      y=cos(30)*15mm,
      z={(0, -sin(30)*15mm)},
    ]
      \def\cylrad{1} 
      \def\cylht{4} 
      \def\phase{0} 
      \draw
        (-\cylrad, \cylht) -- (-\cylrad, 0) --
        plot[smooth, samples=25, variable=\t, domain=180:360]
          ({cos(\t)*\cylrad}, 0, {-sin(\t)*\cylrad}) --
        (\cylrad, \cylht)
        plot[smooth cycle, samples=51, variable=\t, domain=0:360]
          ({cos(\t)*\cylrad}, \cylht, {-sin(\t)*\cylrad})
      ;
      \draw[dashed]
        plot[smooth, samples=9, variable=\t, domain=0:180]
          ({cos(\t)*\cylrad}, 0, {-sin(\t)*\cylrad})
      ;
      \draw[red, thick]
        plot[smooth, samples=25, variable=\t, domain=180:360]
          ({cos(\t)*\cylrad}, {(\t-180)*\cylht/720}, {-sin(\t)*\cylrad})
      ;
      \draw[name path=A, red, thick]
        plot[smooth, samples=25, variable=\t, domain=180:360]
          ({cos(\t)*\cylrad}, {\cylht/2 + (\t-180)*\cylht/720}, {-sin(\t)*\cylrad})
      ;
      \draw[name path=C,red, thick, densely dashed]
        plot[smooth, samples=25, variable=\t, domain=0:180]
          ({cos(\t)*\cylrad}, {\cylht/4 + \t*\cylht/720}, {-sin(\t)*\cylrad})
      ;
      \draw[red, thick, densely dashed]
        plot[smooth, samples=25, variable=\t, domain=0:180]
          ({cos(\t)*\cylrad}, {3*\cylht/4 + \t*\cylht/720}, {-sin(\t)*\cylrad})
      ;
      \draw[name path=B, blue, thick]
        plot[smooth, samples=25, variable=\t, domain=0:180]
        ({cos(\t+\phase)*\cylrad}, {\cylht/4 + \t*\cylht/720}, {sin(\t+\phase)*\cylrad})
      ;
      \draw[blue, thick]
        plot[smooth, samples=25, variable=\t, domain=0:180]
        ({cos(\t+\phase)*\cylrad}, {3*\cylht/4 + \t*\cylht/720}, {sin(\t+\phase)*\cylrad})
      ;
      \draw[blue, thick, densely dashed]
        plot[smooth, samples=25, variable=\t, domain=180:360]
          ({cos(\t+\phase)*\cylrad}, {(\t-180)*\cylht/720}, {sin(\t+\phase)*\cylrad})
      ;
      \draw[name path=D, blue, thick, densely dashed]
        plot[smooth, samples=25, variable=\t, domain=180:360]
          ({cos(\t+\phase)*\cylrad}, {\cylht/2 + (\t-180)*\cylht/720}, {sin(\t+\phase)*\cylrad})
      ;
      \tikzfillbetween[of=A and B]{gray, opacity=0.2}
      ;
      \tikzfillbetween[of=C and D]{gray, opacity=0.2}
      ;
      \node[right] at (\cylrad, 3.1*\cylht/4, 0) {\color{blue} $x^+=\text{const.}$}
      ;
      \node[right] at (\cylrad, 2.8*\cylht/4, 0) {\color{red} $x^-=\text{const.}$}
      ;
      \draw[fill]
      (\cylrad, \cylht/4, 0) circle [radius=1pt] node[right] {$x_0$}
      ;
      \draw[thick, ->] (-1.1*\cylrad,0,0) node[circle,fill,inner sep=1pt,]{}
      -- (-1.1*\cylrad,1,0) node[anchor=north east] {$t$}
      ;
      \draw[thick, ->]
        plot[smooth, samples=25, variable=\t, domain=180:240]
        ({1.1*cos(\t)*\cylrad}, 0, {-sin(\t)*1.2*\cylrad})
        node[anchor=north] {$x$}
      ;
    \end{tikzpicture}
  \end{subfigure}
  \caption{Cylindrical manifold}
  \label{fig:cylinder}
\end{figure}
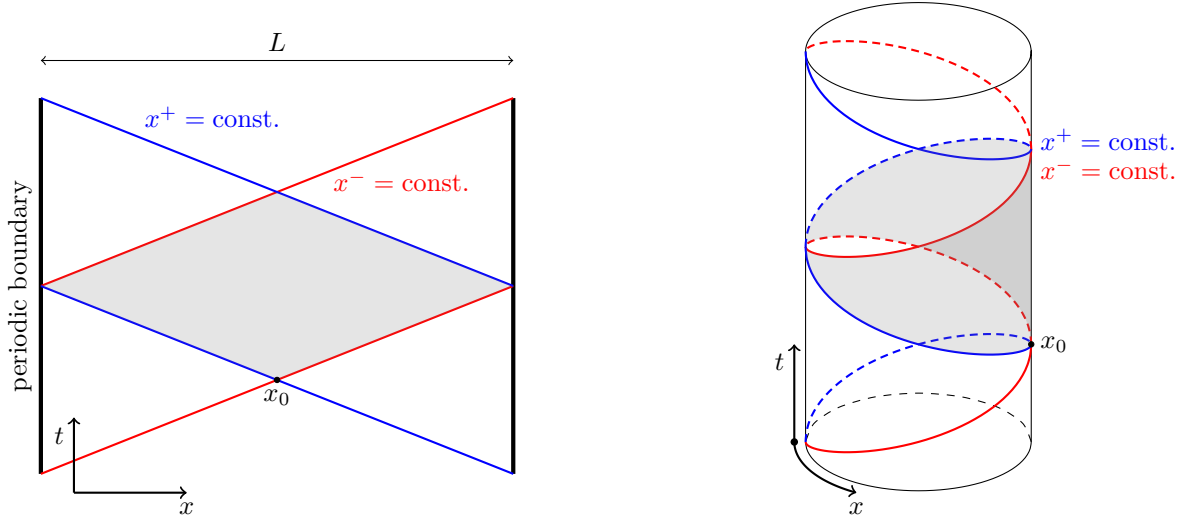

Additionaly, in order to ensure analyticity, the usual choice is to compactify space dimension, in Minkowski space, by setting periodic boundary conditions, \citep{QFTCurvedSpace,CFTDiFrancesco}, i.e. identifying
\begin{equation}
  x\sim x+L,
\end{equation}
where $L$ is the periodicity length. Thus, effectively working with a cylindrical manifold as illustrated by Figure \ref{fig:cylinder}. Moreover, the transformation into the light-cone coordinates leads to coordinates $x^+$ and $x^-$ both to be periodic, with equal periodicity $P$,
\begin{equation}\label{eq:Pperiodicity}
  P =  L/{\sqrt2}.
\end{equation}
This allows the Fourier decomposition of $f^\mu$ into a discrete set of solutions.

Picking a point on the cylinder $x_0 = (x^+_0, x^-_0)$ and drawing the lines $x^+=\text{const.}$ and $x^-=\text{const.}$ starting at the point $x_0$ divides the cylinder into periodic rhombi-like ``tiles'' with one diagonal equal to the spatial periodicity $L$. The shaded regions in Figure \ref{fig:cylinder} illustrate this. This effectively compactifies the $\R\times \mathbb{S}^1$ into $\mathbb{S}^1\times\mathbb{S}^1$. An important remark here is that this compactification onto $\mathbb{S}^1\times\mathbb{S}^1$ may or may not occur, depending on the theory and its inherent existence of $x^\pm$ split.

Another peculiarity of light-cone coordinates is the requirement to set boundary conditions on both lines, $x^+=\text{const.}$ and $x^-=\text{const.}$, passing through the point $x_0$, to find a unique solution. Initial value kind of a problem, where one of the $x^\pm$ coordinates is promoted to be ``time'' and the initial values of fields are given on the line of constant ``time'', is not enough to obtain the full solution.

For the Euclidean metric the plane $\R^2$ is immediately isomorphic to the complex plane $\CC$, for which the natural compactification is the Riemann sphere. Therefore, the whole plane is already in analytic domain. One immediate consequence of dealing with Euclidean space is the loss of a natural choice of time coordinate which plays a special role in Hamiltonian mechanics. Here, all directions are equal, and it is a matter of convenience to promote one coordinate to the special time-like position. Moreover, Euclidean space gives greater freedom to construct coordinates and in two dimensions the most convenient choice is \emph{radial parametrization}, that spans the space by concentric circles around the origin and the radius of the circle represents the ``time'' coordinate $x^0$, while the position at a circle represents the ``space'' coordinate $x^1$. This naturally leads to periodic space dimension. Imposing the periodicity of $x^1$ to be $L$, the corresponding coordinates are $w = e^{\frac{2\pi}{L}z}=e^{\frac{2\pi}{L}(x^0+ix^1)}$, $\bar w = e^{\frac{2\pi}{L}\bar z} = e^{\frac{2\pi}{L}(x^0-ix^1)}$. These coordinates can be viewed as a projection of a cylinder onto a plane. The strength of this coordinate system becomes evident when we realize that integrals over the space domain (e.g. Noether charge \eqref{eq:NoetherCharge}) turn into integrals over closed curves (circles), thus, naturally pointing to the use of residue theorem.

It should be stressed that  $z$ and $\bar z$ are taken as independent, rather than complex conjugated\footnote{Complex conjugation of $z$ is denoted here by $z^*$.} variables. The Euclidean space is tacitly \textit{complexified}, i.e. first the coordinates $x^0$ and $x^1$ are promoted to complex variables, which formally means that we start from the manifold $\CC\times\CC$ rather than $\R\times\R$. In that way transition to the new variables $z$ and $\bar z$ is just a change of variables, henceforth $z$ and $\bar z$ are independent variables, each in its own $\CC$\hyp{}plane. The physical theory is obtained after solving the problem at hand and restricting solutions only to $z^*=\bar z$. It is this cut that turns the conformal Killing equation $\bar\partial f = 0$ into the holomorphic condition for $f$ (similarly, antiholomorphic condition for $\bar f$). Since it is expected to do this cut in the end, it is common practice to restrict to (anti)holomorphic solutions at the very beginning.

The cylindrical manifold and the radial parametrization are very close to each other an it is just a matter of a proper choice of periodicity parameters, and of the Wick rotation, to transform one situation into the other. In fact, solving the problem in $\CC\times\CC$ provides result for Minkowski spacetime if restricted to the cut where $z$ and $\bar z$ are pure imaginary, i.e.  $iz\in \R$ and $i\bar{z} \in\R$. In other words, if ``space'' coordinate $x^1$ is real and ``time'' coordinate $x^0$ is imaginary. That is formally equivalent to Wick-rotating time.

\section{All the Liouville actions used in this paper}\label{app:Liouvilles}

Throughout the paper we discussed Liouville theory at eight separate instances, each time in a different context. To discern, different subscripts for the corresponding
actions\footnote{Except for the flat case, where the letter used to indicate the action is non-curly $A$ instead of $\mathcal A$.} were used. For the sake of clarity, we collect all different definitions here, and explain the nomenclature.

%
%
%
%
%
%
%

Flat spacetime, see equation \eqref{eq:FlatLiouvilleTheory}
{%
\setlength{\belowdisplayskip}{3pt}%
\setlength{\abovedisplayskip}{3pt}%
\begin{align*}
    A_{\scriptscriptstyle{L}}[\Phi] =& \int \dd[2]{x}
    \qty(
      \frac{1}{2}\eta^{\mu\nu}\partial_\mu\Phi\partial_\nu\Phi
      - \frac{m^2}{\beta^2}e^{\beta\Phi}).
  \intertext{\indent \emph{Diffeomorphism} invariant, see equation \eqref{eq:MinimalLiouville}}
    \mathcal A_{\scriptscriptstyle{D}}[\Phi] =& \int \dd[2]{x}\sqrt{-g}
    \qty(
      \frac{1}{2}g^{\mu\nu}\nabla_\mu\Phi\nabla_\nu\Phi
      - \frac{m^2}{\beta^2}e^{\beta\Phi}).
  \intertext{\indent \emph{Non-Minimally} coupled to metric, see equation \eqref{eq:CurvedLiouville}}
    \mathcal A{}_{\scriptscriptstyle{NM}}[\Phi]=&\int\dd[2]{x}\sqrt{-g}
    \qty(
      \frac12g^{\mu\nu}\nabla_\mu\Phi\nabla_\nu\Phi
      - \frac{m^2}{\beta^2}e^{\beta\Phi}
      + \alpha R\Phi
    ).
  \intertext{\indent\emph{Weyl-gauged}, see equation \eqref{eq:WeylGaugedLiouville}}
    \mathcal A_{\scriptscriptstyle{W}}[\Phi,W_\mu]
    =& \int \dd[2]{x}\sqrt{-g}
    \qty(
      \frac{1}{2}g^{\mu\nu}\nabla_\mu\Phi\nabla_\nu\Phi
      - \frac{m^2}{\beta^2}e^{\beta\Phi}
      +\frac{2}{\beta}\Phi\nabla_\mu W^\mu
      +\frac{2}{\beta^2}g^{\mu\nu}W_\mu W_\nu
    ).
  \intertext{\indent Ricci-gauged, i.e., $W^\mu$ traded for the \emph{Ricci} scalar $R$, see equation \eqref{eq:RicciGaugedLiouville}}
    \mathcal A_{\scriptscriptstyle{R}}[\Phi] =& \int \dd[2]{x}\sqrt{-g}
    \qty(
      \frac{1}{2}g^{\mu\nu}\nabla_\mu\Phi\nabla_\nu\Phi
      - \frac{m^2}{\beta^2}e^{\beta\Phi}
      +\frac{1}{\beta}\Phi R
      +\frac{2}{\beta^2}g^{\mu\nu}W_\mu W_\nu),
  \intertext{where
    $$2\nabla_\mu W^\mu = R.$$
  \indent\emph{Liouville} theory commonly used, see equation \eqref{eq:CommonLiouville}}
    \mathcal A_{\scriptscriptstyle{L}}[\Phi]=&\int\dd[2]{x}\sqrt{-g}
    \qty(
      \frac12g^{\mu\nu}\nabla_\mu\Phi\nabla_\nu\Phi
      -\frac{m^2}{\beta^2}e^{\beta\Phi}
      +\frac1\beta R\Phi
    ).
  \intertext{\indent \emph{Polyakov} effective action, see equation \eqref{eq:PolyakovLiouville}}
  \mathcal A_{\scriptscriptstyle{LP}}[\Phi] =&~\int\dd[2]{x}\sqrt{-g}
  \qty(
    \frac12g^{\mu\nu}\nabla_\mu\Phi\nabla_\nu\Phi
    -\frac{m^2}{\beta^2}e^{\beta\Phi}
    +\frac1\beta R\Phi
  )
  \\
  &+\frac1{2\beta^2} \int \dd[2]{x}\dd[2]{y} \sqrt{-g(x)}R(x)K(x,y)\sqrt{-g(y)}R(y),
  \intertext{where
  $$\nabla^2_x K(x,y) = \frac{1}{\sqrt{-g(x)}}\delta^{(2)}(x-y).$$
  \indent\emph{Liouville} action with the \emph{Deser-Jackiw} improvement, see equation \eqref{eq:DeserJackiwLiouville}}
    \mathcal A_{\scriptscriptstyle{LDJ}}[\Phi] =& \int \dd[2]{x}\sqrt{-g}
    \qty(
      \frac{1}{2}g^{\mu\nu}\nabla_\mu\Phi\nabla_\nu\Phi
      - \frac{m^2}{\beta^2}e^{\beta\Phi}
      +\frac{1}{\beta}\Phi R
      +\frac{2}{\beta^2}g_{\mu\nu}W^\mu_{\scriptscriptstyle{DJ}} W^\nu_{\scriptscriptstyle{DJ}}),
  \intertext{where
  $$W^\mu_{\scriptscriptstyle{DJ}} =
  \frac{\varepsilon^{\mu\nu}}{2\sqrt{-g}}
  \qty[
    \frac{\varepsilon^{\alpha\beta}}{\sqrt{-g}}
    \Gamma_{\beta\alpha\nu} 
    +
    (\cosh\sigma -1)\partial_\nu\gamma
    +
    \partial_\nu r
  ].$$}
\end{align*}}

\section{On Vielbein and spin-connection}\label{app:Vielbeins}
The provided solution $W^\mu_{\scriptscriptstyle{DJ}}$ for
\begin{equation*}
  2\nabla_\mu W^\mu = R\,,
\end{equation*}
can be written explicitly without introducing the parametrization of the metric, but rather introducing Vielbein formalism, \citep{DeserJackiw96,Iorio2011WeylGaugeGraphene}. Such reformulation of the problem in two dimensions shows explicitly that $W^\mu_{\scriptscriptstyle{DJ}}$ is proportional to the spin connection, which is demonstrated in the following.

In order to relate flat and curved spaces, one way is to introduce local ``orthonormal'' bases, defined on the tangent space at each point of the curved manifold. The transformation between the local orthonormal basis, labeled by ``flat'' Latin indices $a,b,c,\ldots$, and coordinate basis, labeled by ``curved'' Greek indices $\mu,\nu,\lambda,\ldots$, is realized by invertible matrices $e^a_\mu$, called vielbein\footnote{
  The word ``viel'' is often replaced by German word characterizing the dimension number, e.g. zweibein for two dimensions, vierbein for four dimensions. Some authors prefer the names tetrads, orthonormal basis or frame fields.
} defined by
\begin{equation}\label{eq:VielbeinDefinition}
  g_{\mu\nu} = e^a_\mu \eta_{ab}e^b_\nu\,.
\end{equation}
Introducing the inverse $E^\mu_a$
\begin{equation}
  e^a_\mu E_a^\nu = \delta_\mu^\nu\,,
  \qquad
  e^a_\mu E_b^\mu = \delta_b^a\,,
\end{equation}
the \eqref{eq:VielbeinDefinition} can be expressed as
\begin{equation}
  E^\mu_a g_{\mu\nu}E^\nu_b = \eta_{ab}\,.
\end{equation}

A vector field $V$, at any given point can then be written either in the coordinate basis with vector components $V^\mu$, or in local flat basis with components $V^a$. The two are connected by
\begin{equation}
  V^\mu = E^\mu_a V^a\,,\qquad
  V^a = e^a_\mu V^\mu\,.
\end{equation}
While the Greek indices respond to general coordinate transformations
\begin{equation*}
  V'^\mu = \pdv{x'^\mu}{x^\nu} V^\nu\,,
\end{equation*}
the Latin indices only allow local Lorentz transformations,
\begin{equation*}
  V'^a = \Lambda^a{}_bV^b\,.
\end{equation*}
Different transformations would not preserve the metric $\eta_{ab}$ and the local orthonormal basis.

The above can easily be generalized to tensors of any rank, \citep{CarrollRelativity}. Moreover, multiindex objects are well-defined and can undergo simultaneous transformations in both kinds of indices, e.g. vielbein transform
\begin{equation}
  e'^a_\mu = \Lambda^a{}_b\pdv{x^\nu}{x'^\mu}e^b_\nu\,.
\end{equation}

Let us first here discuss a bit more in detail the implications of the Vielbein formalism. This was a simple introduction of a local change of basis. The important question is then how tensors in the local basis are parallel transported. Or equivalently, how the covariant derivative looks like. Just like in the coordinate basis the introduction of (Levi-Civita) connection, $\Gamma^\lambda{}_{\mu\nu}$, cancels the non-tensorial contribution of the partial derivative, another connection, $s_\mu{}^a{}_b$, is introduced here to take care of Latin flat indices. This is called \emph{spin connection}\footnote{
  The main use of vielbein and spin connection is for dealing with spinors, \citep{QFTCurvedSpace}. From here the name.
} and the covariant derivative of a $(1, 1)$ tensor $T^a{}_b$ in the local basis is then
\begin{equation}
  \nabla_\mu T^a{}_b
  \equiv
  \partial_\mu T^a{}_b
  + s_\mu{}^a{}_c T^c{}_b
  - s_\mu{}^c{}_b T^a{}_c\,.
\end{equation}
Again, the multiindex case is possible and both kinds of connections may appear in the covariant derivative. Moreover, assuming the metric compatibility, $\nabla_\mu g_{\alpha\beta} = 0$, and consequently $\nabla_\mu \eta_{ab}=0$, it follows that the spin connection is antisymmetric in the Latin indices
\begin{equation}
  s_{\mu ab} = - s_{\mu ba}\,.
\end{equation}

A special example of multiindex covariant derivative is the one of vielbein, for which the so-called full covariant derivative vanish by construction (the ``Vielbein postulate''),
\begin{equation}
  \nabla_\mu e^a_\nu = \partial_\mu e^a_\nu - {\Gamma^\lambda}_{\mu\nu} e^a_\lambda
  + {{s_\mu}^a}_b e^b_\nu = 0\,.
\end{equation}
This equation can also serve as a starting point for the reformulation of the Riemann tensor in terms of spin connection, rather than Levi-Civita connection, which can be written as
\begin{equation}\label{eq:ChristoffelSpinConnection}
  {\Gamma^\lambda}_{\mu\nu} = E^\lambda_a(\partial_\mu e_\nu^a + {{s_\mu}^a}_b e^b_\nu)\,.
\end{equation}
Thus, the Riemann tensor
\begin{equation*}
  {R^\rho}_{\sigma\mu\nu} \equiv
  \partial_\mu {\Gamma^\rho}_{\nu\sigma}
  - \partial_\nu {\Gamma^\rho}_{\mu\sigma}
  + {\Gamma^\rho}_{\mu\lambda}{\Gamma^\lambda}_{\nu\sigma}
  - {\Gamma^\rho}_{\nu\lambda}{\Gamma^\lambda}_{\mu\sigma}\,,
\end{equation*}
after substitution of \eqref{eq:ChristoffelSpinConnection} becomes
\begin{equation}\label{eq:RiemannSpinConnection}
  \begin{split}
    {R^\lambda}_{\rho\mu\nu} =&~ E^\lambda_a e^b_\rho
    \qty(
      \partial_\mu {{s_\nu}^a}_b - \partial_\nu {{s_\mu}^a}_b
      +{{s_\mu}^a}_c{{s_\nu}^c}_b - {{s_\nu}^a}_c{{s_\mu}^c}_b
    )\,,
    \\
    =&~ E^\lambda_a e^b_\rho {R^a}_{b\mu\nu}\,.
  \end{split}
\end{equation}

The EMT of a theory formulated in Vielbein formalism, obtained by variation of an action $\mathcal A$, is
\begin{equation}
  T^{\mu\nu} = -\frac{1}{2e}\qty(
    \fdv{\mathcal A}{e^a_\mu}\eta^{ab}E^\nu_b
    + \fdv{\mathcal A}{e^a_\nu}\eta^{ab}E^\mu_b
  )\,,
\end{equation}
where
\begin{equation}
  e\equiv \det(e^a_\mu) = \sqrt{-g}\,.
\end{equation}

\subsubsection*{Two dimensions}
In two dimensions the Levi-Civita symbol has only two indices, $\varepsilon^{ab}$, and any antisymmetric object has to be proportional the Levi-Civita symbol, since in two dimensions it itself spans the whole space of antisymmetric objects. Thus, in the case of metric compatible spin connection, it is expected
\begin{equation}
  s_{\mu ab}\propto s_\mu \varepsilon_{ab}\,.
\end{equation}
Defining $s_\mu$ as
\begin{equation}
  s_\mu \equiv s_{\mu ab}\varepsilon^{ab}\,,
\end{equation}
it immediately follows
\begin{equation}
  s_{\mu ab} = -\frac12 s_\mu \varepsilon_{ab}\,.
\end{equation}
It should be stressed that although $s_\mu$ does not have any latin indices it is \emph{not} a scalar under local Lorentz transformations.

Using the previous result together with
\begin{align}
  \varepsilon^{ab}\varepsilon_{b}{}^c =&~ \eta^{ac}\,,
  \\
  E^\mu_a\varepsilon^{ab}E^\nu_b =&~ \frac{\varepsilon^{\mu\nu}}{\sqrt{-g}}\,,
\end{align}
leads to great simplification of the Riemann tensor \eqref{eq:RiemannSpinConnection}
\begin{equation}
  {R^\lambda}_{\rho\mu\nu} = \frac{\varepsilon^{\lambda\sigma}}{\sqrt{-g}}g_{\sigma\rho}\frac12\qty(\partial_\nu s_\mu - \partial_\mu s_\nu)\,.
\end{equation}
This, in turn, leads to a simple form of the Ricci scalar
\begin{equation}
  R=\frac{\varepsilon^{\mu\nu}}{\sqrt{-g}}\partial_\nu s_\mu\,,
\end{equation}
which greatly resembles the identification \eqref{eq:WeylRicci2DIdentification}
\begin{equation}
  2\nabla_\mu W^\mu
  =
  \frac2{\sqrt{-g}}\partial_\mu \qty(\sqrt{-g}W^\mu)
  =
  R\,.
\end{equation}

Thus, another form of the local solution to \eqref{eq:WeylRicci2DIdentification} is
\begin{equation}
  W^\mu = - \frac{\varepsilon^{\mu\nu}}{2\sqrt{-g}}s_\nu\,.
\end{equation}
Since $W^\mu$ is proportional to the spin connection this shows once more that a local solution cannot be a vector.

\end{document}